\documentclass[twocolumn, switch]{article} 

\usepackage{preprint}

\usepackage{amsmath, amsthm, amssymb, amsfonts}

\usepackage[numbers,square]{natbib}
\usepackage[utf8]{inputenc}	
\usepackage[T1]{fontenc}	
\usepackage[colorlinks = true,
            linkcolor = purple,
            urlcolor  = blue,
            citecolor = cyan,
            anchorcolor = black]{hyperref}	
\usepackage{booktabs} 		
\usepackage{nicefrac}		
\usepackage{microtype}		
\usepackage{lineno}		
\usepackage{float}			

\usepackage{amsmath}
\usepackage{hyperref}
\usepackage{subcaption}
\usepackage{bbold}
\usepackage{dirtytalk}
\usepackage{multirow}
\usepackage{physics}

\usepackage{newfloat}
\DeclareFloatingEnvironment[name={Supplementary Figure}]{suppfigure}
\usepackage{sidecap}
\sidecaptionvpos{figure}{c}

\usepackage{titlesec}
\titlespacing\section{0pt}{12pt plus 3pt minus 3pt}{1pt plus 1pt minus 1pt}
\titlespacing\subsection{0pt}{10pt plus 3pt minus 3pt}{1pt plus 1pt minus 1pt}
\titlespacing\subsubsection{0pt}{8pt plus 3pt minus 3pt}{1pt plus 1pt minus 1pt}

\usepackage{tikz,xcolor,hyperref}

\definecolor{lime}{HTML}{A6CE39}
\DeclareRobustCommand{\orcidicon}{
	\begin{tikzpicture}
	\draw[lime, fill=lime] (0,0) 
	circle [radius=0.16] 
	node[white] {{\fontfamily{qag}\selectfont \tiny ID}};
	\draw[white, fill=white] (-0.0625,0.095) 
	circle [radius=0.007];
	\end{tikzpicture}
	\hspace{-2mm}
}
\foreach \x in {A, ..., Z}{\expandafter\xdef\csname orcid\x\endcsname{\noexpand\href{https://orcid.org/\csname orcidauthor\x\endcsname}
			{\noexpand\orcidicon}}
}

\title{Enhanced Fuzzy Decomposition of Sound Into Sines, Transients and Noise}

\usepackage{xwatermark}
\newwatermark[firstpage,color=gray!90,angle=0,scale=0.28, xpos=0in,ypos=-5in]{Submitted for publication to the Journal of Audio Engineering Society on October 20th, 2022.}

\usepackage{authblk}

\author{Leonardo Fierro and Vesa Välimäki\\ \texttt{leonardo.fierro@aalto.fi | vesa.valimaki@aalto.fi}}

\affil{Acoustics Lab, Department of Signal Processing and Acoustics, Aalto University, Espoo, Finland}

\begin{document}

\twocolumn[ 
  \begin{@twocolumnfalse} 
  
\maketitle

\begin{abstract}
The decomposition of sounds into sines, transients, and noise is a long--standing research problem in audio processing. The current solutions for this three--way separation detect either horizontal and vertical structures or anisotropy and orientations in the spectrogram to identify the properties of each spectral bin and classify it as sinusoidal, transient, or noise. This paper proposes an enhanced three--way decomposition method based on fuzzy logic, enabling soft masking while preserving the perfect reconstruction property. The proposed method allows each spectral bin to simultaneously belong to two classes, sine and noise or transient and noise. Results of a subjective listening test against three other techniques are reported, showing that the proposed decomposition yields a better or comparable quality. The main improvement appears in transient separation, which enjoys little or no loss of energy or leakage from the other components and performs well for test signals presenting strong transients. The audio quality of the separation is shown to depend on the complexity of the input signal for all tested methods. The proposed method helps improve the quality of various audio processing applications. A successful implementation over a state-of-the-art time-scale modification method is reported as an example. 
\end{abstract}
\vspace{0.35cm}

  \end{@twocolumnfalse} 
] 


\section{INTRODUCTION}
\label{sec:intro}

Decomposing an audio signal into its sinusoidal, transient, and noise (STN) components has been drawing research interest for over two decades \cite{Verma2000, driedger2014extending, fug2016harmonic, damskagg2017audio}. It is a widely used tool in a variety of audio processing applications, ranging from beat tracking \cite{gkiokas2012music} and tonality estimation \cite{faraldo2018tonality} to reduction of spectral complexity in cochlear implants \cite{Lentz2020} and to virtual bass enhancement \cite{moliner2020virtual}. The STN separation is also helpful in time-scale modification \cite{Verma2000, nsabimana2008audio, driedger2013improving}, where it has been combined with the notion of fuzzy logic in order to improve \cite{damskagg2017audio,roberts2019time} or evaluate the audio quality \cite{fierro2020towards}. In all these audio applications, it is helpful to process sine, transient, and noise components independently of each other. This paper proposes improvements to the fuzzy STN decomposition of audio signals. 



The STN separation relies on the assumption that any audio signal can be described as a linear combination of three independent actors: tonal content (sines), impulsive events (transients), and a residual part (noise) that does not belong to either one of the other two classes and adds nuance to the sound. Historically, additive synthesis modeled any sound as a sum of sinusoidal components \cite{Beauchamp1966, Risset1969, Moorer1977}. Serra and Smith expanded the additive synthesis method by introducing the noise class, which was obtained as a residual after a sinusoidal model was subtracted from the original signal \cite{Serra1990}. In the resulting method---called spectral modeling synthesis \cite{Serra1990}---the frequency, amplitude, and phase of the sinusoidal components were estimated from the short-time Fourier transform (STFT) using a method similar to the McAulay-Quatieri algorithm \cite{McAulay1986}. 

The three-way decomposition was first introduced by Verma et al. \cite{Verma2000, verma1997, Verma}, who showed that including a third component for transients was greatly beneficial in the context of signal analysis and synthesis, as it avoided the smearing of transients, which was a weakness in sines + noise models. Levine and Smith also showed that the adaptiveness of the STN model made it suitable for audio compression and for pitch-- and time--scale modification \cite{levine1998sines+}.

Fitzgerald discovered that it was possible to decompose an audio signal into its sinusoidal and transient components by using spectral masks extracted via horizontal and vertical median filtering of the STFT \cite{fitzgerald2010harmonic}. Driedger et al.~\cite{driedger2014extending} reintroduced the three-way separation by updating Fitzgerald's method: the noise component could be obtained by retrieving spurious information after extracting the other two components with median filtering. 

Füg et al. \cite{fug2016harmonic} proposed a follow-up method involving the use of structure tensors (ST) to find predominant orientation angles and anisotropy in the time-frequency signal representation, showing an improvement in the separation quality for sounds with vibrato. 

Other recent approaches for sines--transients separation include a kernel additive matrix \cite{fitzgerald2014harmonic}, non-negative matrix factorization \cite{canadas2017harmonic}, improved sinusoidal modeling \cite{neri2018fast, masuyama2019phase}, and neural networks \cite{drossos2018harmonic}. However, these methods do not involve a third class for the noise component, hence they are not discussed further in this paper.

It should be noted that the STN decomposition does not directly relate to traditional source separation, which usually aims at retrieving musical instruments in a sound mixture \cite{rafii2018overview} or speech from noisy background sources \cite{makino2007blind}. According to the STN formulation, even strongly percussive sources, such as drums, will have a sinusoidal and noise component---unless they are perfect, synthetic pulses---and, similarly, strongly-harmonical sources, such as the violin, will hold, in addition to a sinusoidal part, a transient component in their attack and a noise component to describe the nuances, such as the bowing noise.

While both Driedger et al.~\cite{driedger2014extending} and Füg et al.~\cite{fug2016harmonic} applied hard binary masks to define the sinusoidal, transient, and noise classes, Damskägg and Välimäki \cite{damskagg2017audio} introduced the concept of fuzzy logic in the context of time-scale modification. The fuzzy classification (FZ) allows spectral bins to simultaneously contribute to the three classes, providing a more refined basis for the three-way separation \cite{damskagg2017audio}. This decomposition method was then extended to objective evaluation by Fierro and Välimäki \cite{fierro2020towards} and improved by Moliner et al.~\cite{moliner2020virtual} to allow perfect reconstruction by ensuring that the three soft spectral masks sum up to unity. 


This work proposes a novel way to estimate fuzzy soft masks for STN decomposition of audio signals. The proposed method allows for intermediate classifications of the spectral bins between two components---sines v.~noise and transients v.~noise. This two-stage decomposition is shown to improve the overall sound quality of the separated components, in particular for transients. 
The masks ensure perfect reconstruction and are optimized for each class to have a large constant region followed by a fast but smooth transition to the adjacent class. The transition slope is refined for both decomposition stages to provide the best separation quality.

The rest of this paper is structured as follows. Sec.~\ref{sec:separation} discusses previous three-way separation techniques. Sec.~\ref{sec:method} introduces the new STN decomposition method, which optimally extracts the sinusoidal and transient components. Sec.~\ref{sec:evaluation} evaluates the proposed method against three previous techniques. Sec.~\ref{sec:TSM} applies the proposed method to time-scale modification, and Sec.~\ref{sec:conclusion} concludes.

\section{RELATED WORK}
\label{sec:separation}


This section summarizes three previous STN decomposition methods based on a spectrogram representation of the input signal. A spectrogram ${X}$ is a $K$-by-$M$ matrix representing the time-frequency behavior of audio signal $x$. Each element $X(m, k)$ is computed using the STFT: 
\begin{equation}
    X(m,k) = \sum_{n = 0}^{L-1} x(n+m H)\ w(n)\ e^{-j\omega_kn},
\end{equation}
\noindent where $n$ is the sample index, $m = 0, 1, 2, ...\ M-1$ is the temporal frame index, $k = 0, 1, 2, ...\ K-1$ is the spectral bin index, $w$ is the analysis window, $H$ is the hop size, $L$ is the window length in samples, which is assumed to be even, $j$ is the imaginary unit, and $\omega_k$ is the normalized central frequency of the $k^{\text{th}}$ spectral bin. 

\begin{figure}[t!]
\center
\includegraphics[width=0.75\columnwidth]{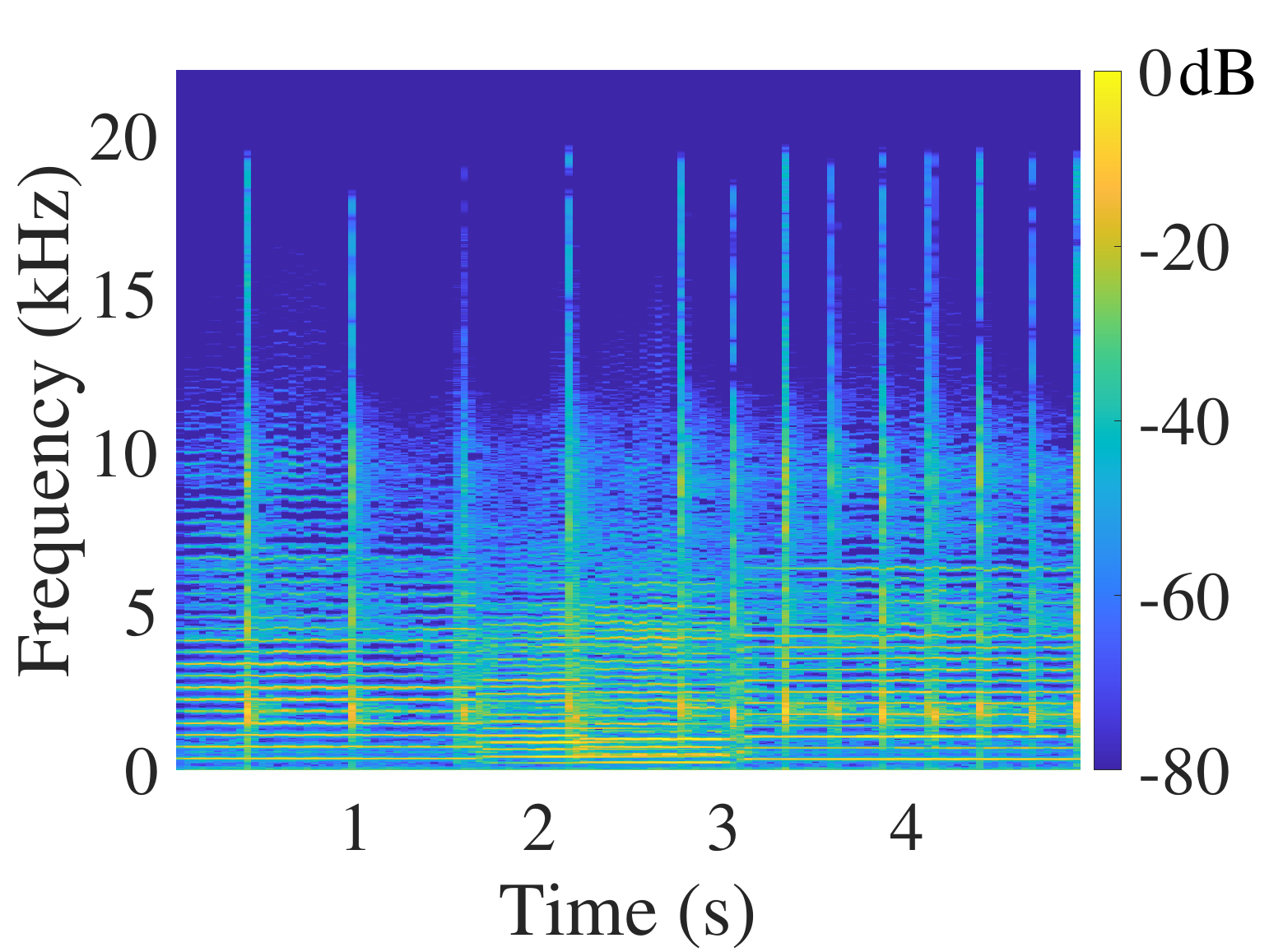}
\caption{\label{fig:Orig}{Spectrogram of a test signal consisting of the castanets and the violin playing simultaneously.}}
\end{figure}

\subsection{Harmonic--Percussive--Residual Separation}

The Harmonic--Percussive--Residual (HPR) separation \cite{driedger2014extending} builds upon the Harmonic--Percussive (HP) method for sines--transients decomposition \cite{fitzgerald2010harmonic}. Fitzgerald noted that since sinusoids form flat lines in time direction in the spectrogram and, vice versa, impulsive events appear as flat lines in the frequency direction, they can be detected (suppressed) using a median filter \cite{fitzgerald2010harmonic}. Fig.~\ref{fig:Orig} shows the spectrogram of a signal  consisting of a mixture of violin and castanets, whose time-- and frequency--direction ridges are noticeable.

Horizontal (time-oriented) and vertical (frequency-oriented) median filtering can be applied to the spectrogram $X(m,k)$ to highlight the desired component and suppress the other \cite{fitzgerald2010harmonic}:
\begin{multline}
X_\textrm{h}(m,k) \\
= \textrm{med}\Big[|X(m-\frac{L_\textrm{h}}{2}+1,k)|,...,|X(m+\frac{L_\textrm{h}}{2},k)|\Big]
\end{multline}
and
\begin{multline}
X_\textrm{v}(m,k) \\ 
= \textrm{med}\Big[|X(m,k-\frac{L_\textrm{v}}{2}+1)|,...,|X(m,k+\frac{L_\textrm{v}}{2})|\Big],
\end{multline}

\noindent where $\textrm{med}[\cdot]$ is the median function, and $X_\textrm{h}$ and $X_\textrm{v}$ are the resulting horizontally- and vertically-enhanced magnitude spectrograms, respectively. Parameters $L_\textrm{h}$ and $L_\textrm{v}$ are the median filter lengths (in samples) in the time and frequency directions, respectively. 

Matrices $X_\textrm{h}$ and $X_\textrm{v}$ are then used to extract the tonalness $R_\textrm{s}$ and transientness $R_\textrm{t}$ matrices with the following elements \cite{fitzgerald2010harmonic}:

\begin{equation} \label{eq:tonalness}
    R_\textrm{s}(m,k) = \frac{X_\textrm{h}(m,k)}{X_\textrm{h}(m,k)+X_\textrm{v}(m,k)}
\end{equation}
and
\begin{equation} \label{eq:transientness}
    R_\textrm{t}(m,k) = 1-R_\textrm{s}(m,k) = \frac{X_\textrm{v}(m,k)}{X_\textrm{h}(m,k)+X_\textrm{v}(m,k)},
\end{equation}

\noindent respectively. Fitzgerald \cite{fitzgerald2010harmonic} used $R_\textrm{s}$ and $R_\textrm{t}$ directly as spectral masks, whereas Driedger et al.~\cite{driedger2014extending} later introduced a controllable separation factor $\beta$ and a third class (noise) to describe those parts of the sound that are neither sines nor transients. 

From Eqs.~\eqref{eq:tonalness} and \eqref{eq:transientness}, a set of hard spectral masks $S$ (sinusoidal), $T$ (transient), and $N$ (noise) can be derived as follows \cite{driedger2014extending}:
\begin{equation}
S(m,k) = \begin{cases} 1, & \mbox{if } R_\textrm{s}(m,k) / R_\textrm{t}(m,k) > \beta \\ 0, & \mbox{otherwise,}\end{cases}
\end{equation}
\begin{equation}
T(m,k) = \begin{cases} 1, & \mbox{if } R_\textrm{t}(m,k) / R_\textrm{s}(m,k) > \beta \\ 0, & \mbox{otherwise,}\end{cases}
\end{equation}
and
\begin{equation} \label{eq:noise}
    N(m,k) = 1 - S(m,k) - T(m,k).
\end{equation}

\begin{figure}[t!]
\center
\includegraphics[width=0.86\columnwidth]{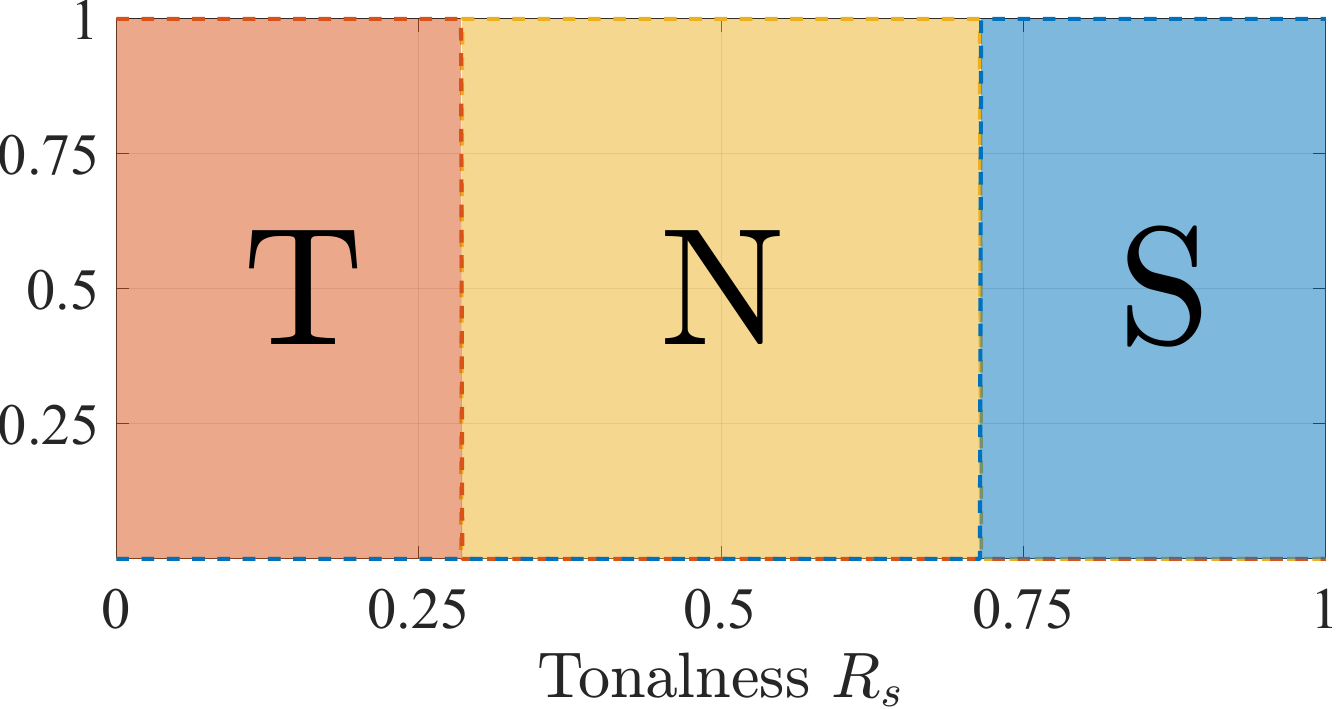}
\caption{\label{fig:HPR}{Hard masks for transients, noise, and sines, as used in the HPR method \cite{driedger2014extending}, for separation factor $\beta = 2.5$.}}
\end{figure}

\noindent Their relationship for a chosen $\beta$ is shown in Fig.~\ref{fig:HPR}. The spectral masks are then imposed on $X(m,k)$  
to retrieve the three desired spectral components:
\begin{equation} \label{eq:masks}
    X_\textrm{s} =  S\odot  X, \qquad X_\textrm{t} =  T\odot  X, \qquad X_\textrm{n} =  N\odot  X,
\end{equation}

\noindent where $\odot$ represents the Hadamard product, or element-wise multiplication.

It has been observed that the quality of the HPR separation largely varies for the sinusoidal and the transient components depending on the choice of the analysis window length $L$ \cite{bonada2000automatic,tachibana2013singing,driedger2014extending}. 
A large window length $L$ for the STFT, ensuring sufficient frequency resolution but poor time resolution, results in a faultless extraction of sines but a low-quality transient output; conversely, a smaller value of $L$ leads to a better extraction of the transient component but a worse description of sines. 

\begin{figure*}[t!]
\center
\begin{subfigure}[t]{0.32\textwidth}
\centering
\includegraphics[width=\textwidth]{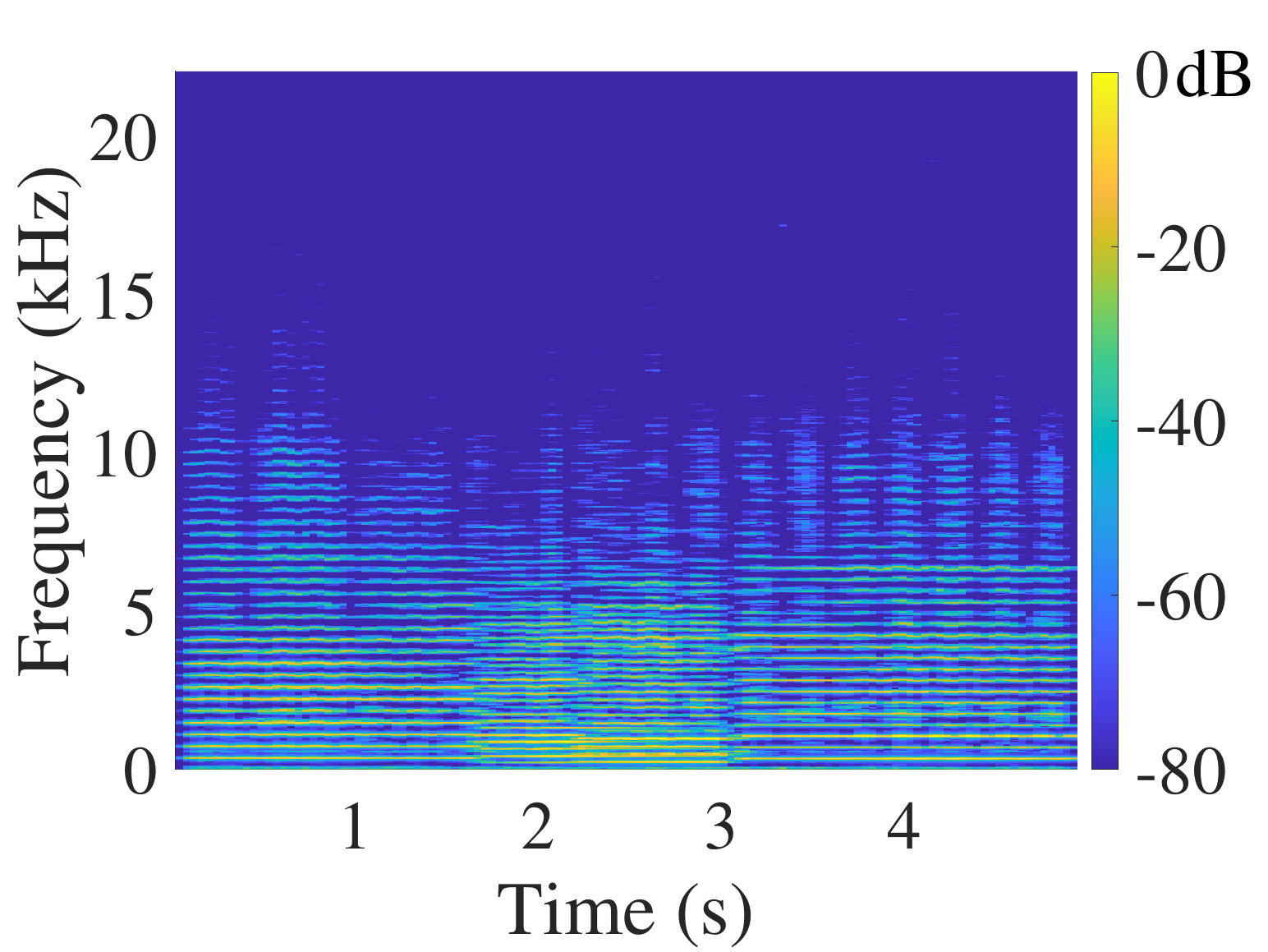}
\caption{Sines}
\end{subfigure}
\begin{subfigure}[t]{0.32\textwidth}
\centering
\includegraphics[width=\textwidth]{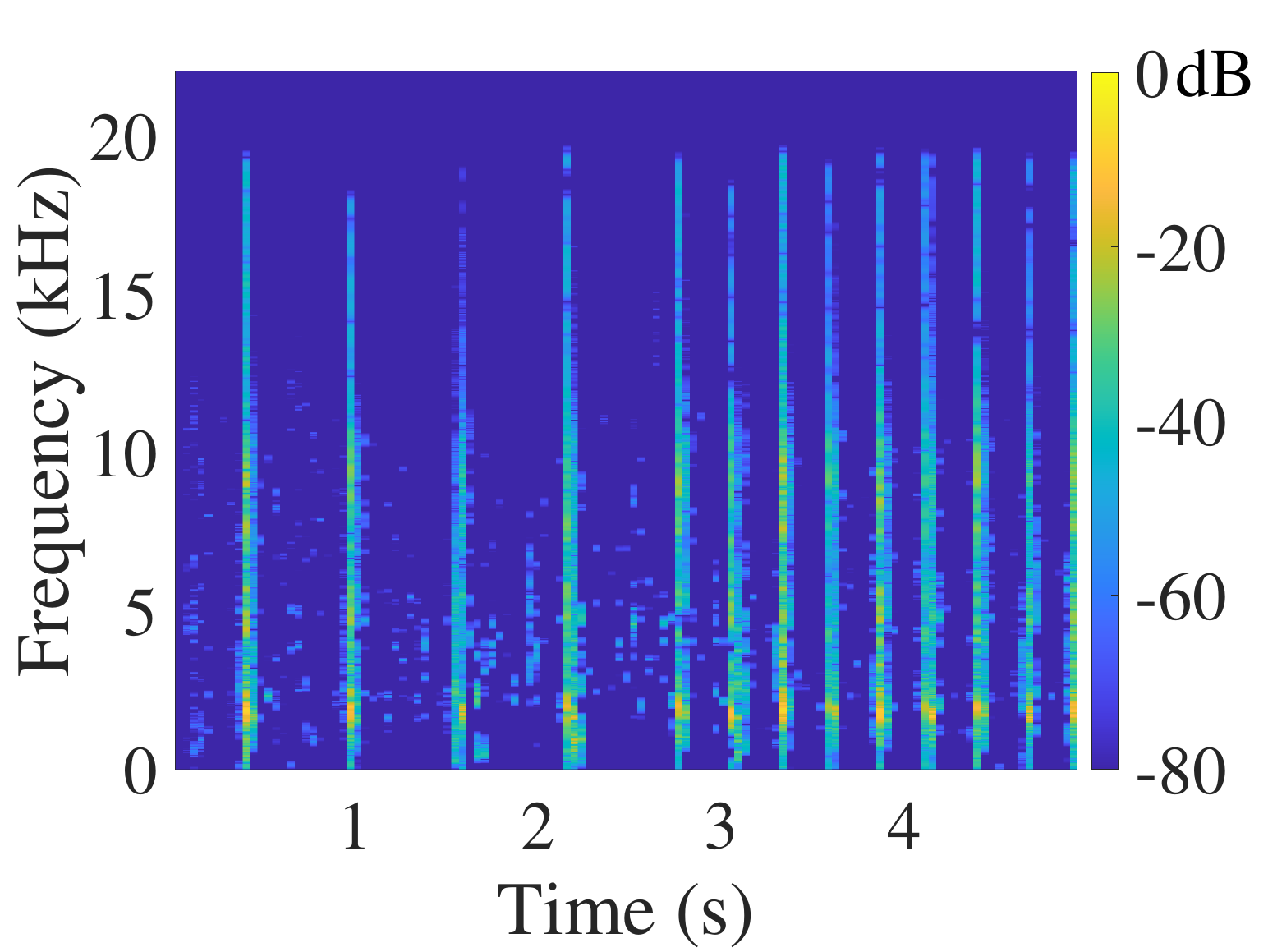}
\caption{Transients}
\end{subfigure}
\begin{subfigure}[t]{0.32\textwidth}
\centering
\includegraphics[width=\textwidth]{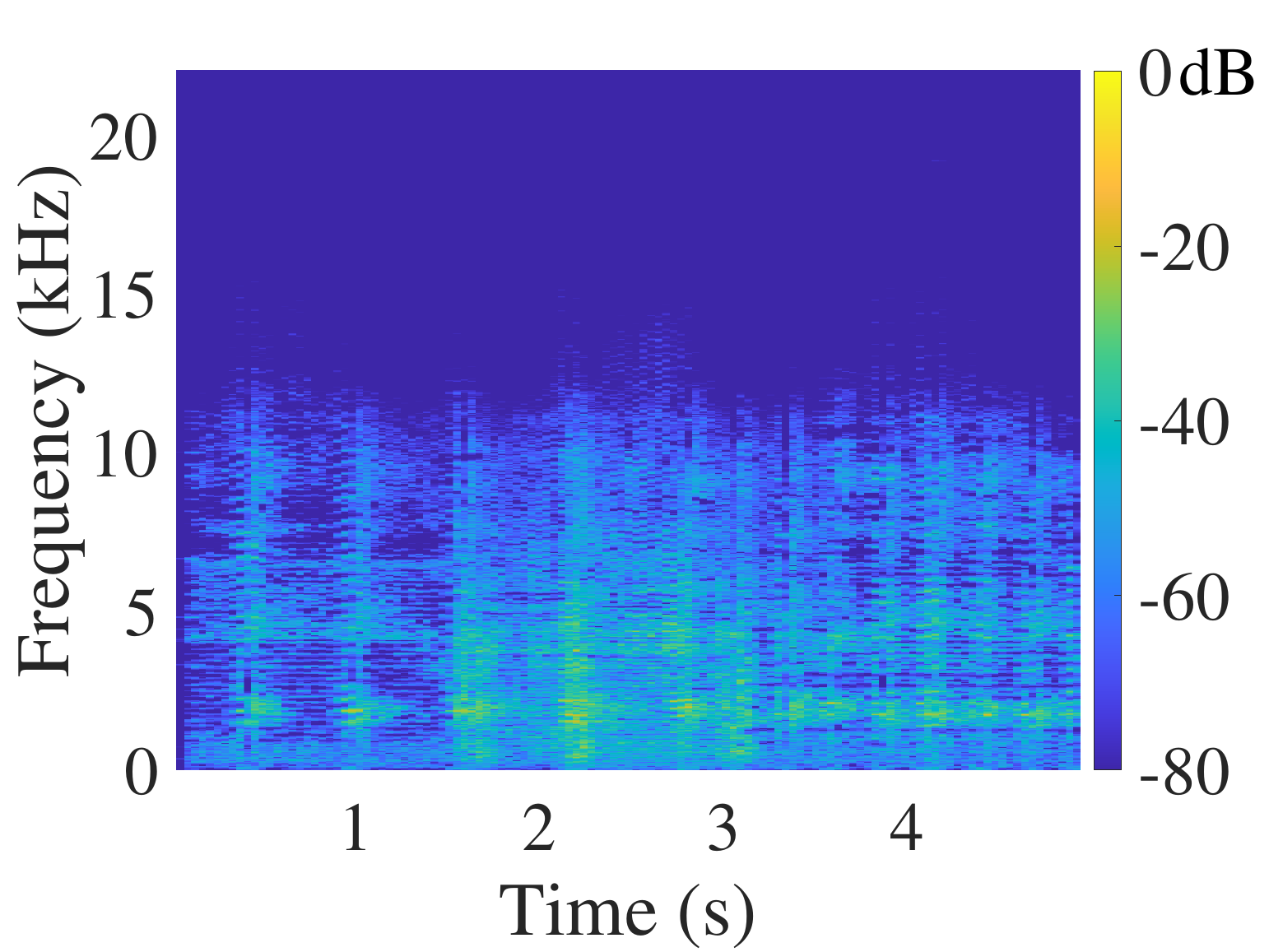}
\caption{Noise}
\end{subfigure}
\caption{\label{fig:STFTHPR} STN separation performed over the mixture of castanets and violin using HPR method.}
\end{figure*}

To overcome the time-frequency limitation, Driedger et al.~\cite{driedger2014extending} divided the decomposition process can be divided into two cascaded iterations \cite{driedger2014extending}. In the first stage, a longer analysis window is applied to extract the sinusoidal component \cite{driedger2014extending}, while transients and noise remain mixed together:
\begin{equation} \label{eq:firstround}
    x_\textrm{s} =  \text{ISTFT}\Big[S_1\odot  X\Big], 
\end{equation}\begin{equation} \label{eq:secondround}
    x_\textrm{res} =  \text{ISTFT}\Big[(T_1 + N_1)\odot  X \Big],
\end{equation}

\noindent where ISTFT is the Inverse STFT. Subsequently, the residual from the first stage is separated again with shorter windowing, leading to the final decomposition \cite{driedger2014extending}:
\begin{equation} \label{eq:secondround}
    x_\textrm{t} =  \text{ISTFT}\Big[T_2\odot  X_\textrm{res}\Big],
\end{equation}\begin{equation} \label{eq:secondround}
    x_\textrm{n} =  \text{ISTFT}\Big[(S_2 + N_2)\odot  X_\textrm{res} \Big].
\end{equation}
\noindent
The noise signal $x_\textrm{n}$ will also contain residuals of sine components, unless they were perfectly separated on the first stage. Fig.~\ref{fig:STFTHPR} shows the separated STN components of the example audio signal used above obtained with the HPR method.

\subsection{STN Separation Based on Structure Tensor}

Füg et al.~\cite{fug2016harmonic} noted that sounds exhibiting vibrato, which carry tonal information and are perceived as sines, do not present strictly horizontal structures in the spectrogram. This results in a leakage of energy in-between different spectral components. The strictness of the median filtering can be overcome using a structure tensor, a widely used tool in image processing, to obtain a measure of the frequency change rate and local anisotropies in the spectrogram, which will then be used as features to define the spectral masks \cite{fug2016harmonic}.

The structure tensor matrix is obtained from the partial derivatives of the spectrogram with respect to time and frequency, and the orientation angles $\alpha$ and the anisotropy $C$ of the spectral bins are computed from the eigenvalues and the eigenvectors of such a matrix, as described in \cite{fug2016harmonic}. The instantaneous frequency change rate $R$ is computed for each bin from the orientation angles:
\begin{equation}\label{eq:alpha}
    R(m,k) = \frac{{f_\textrm{s}}^2}{HM} \tan{[\alpha(m,k)]},
\end{equation}

\noindent where $f_\textrm{s}$ is the sample rate. The spectral masks are then obtained as follows: 
\begin{equation}
S(m,k) = 
\begin{cases} 
1, & \mbox{if } |R(m,k)|\leq r_\textrm{s}\ \wedge\ C(m,k) > c \\ 
0, & \mbox{otherwise,}
\end{cases}
\end{equation}
\begin{equation}
T(m,k) = 
\begin{cases} 
1, & \mbox{if } |R(m,k)|\geq r_\textrm{t}\ \wedge\ C(m,k) > c \\ 
0, & \mbox{otherwise,}
\end{cases}
\end{equation}

\begin{figure*}[t!]
\center
\begin{subfigure}[t]{0.32\textwidth}
\centering
\includegraphics[width=\textwidth]{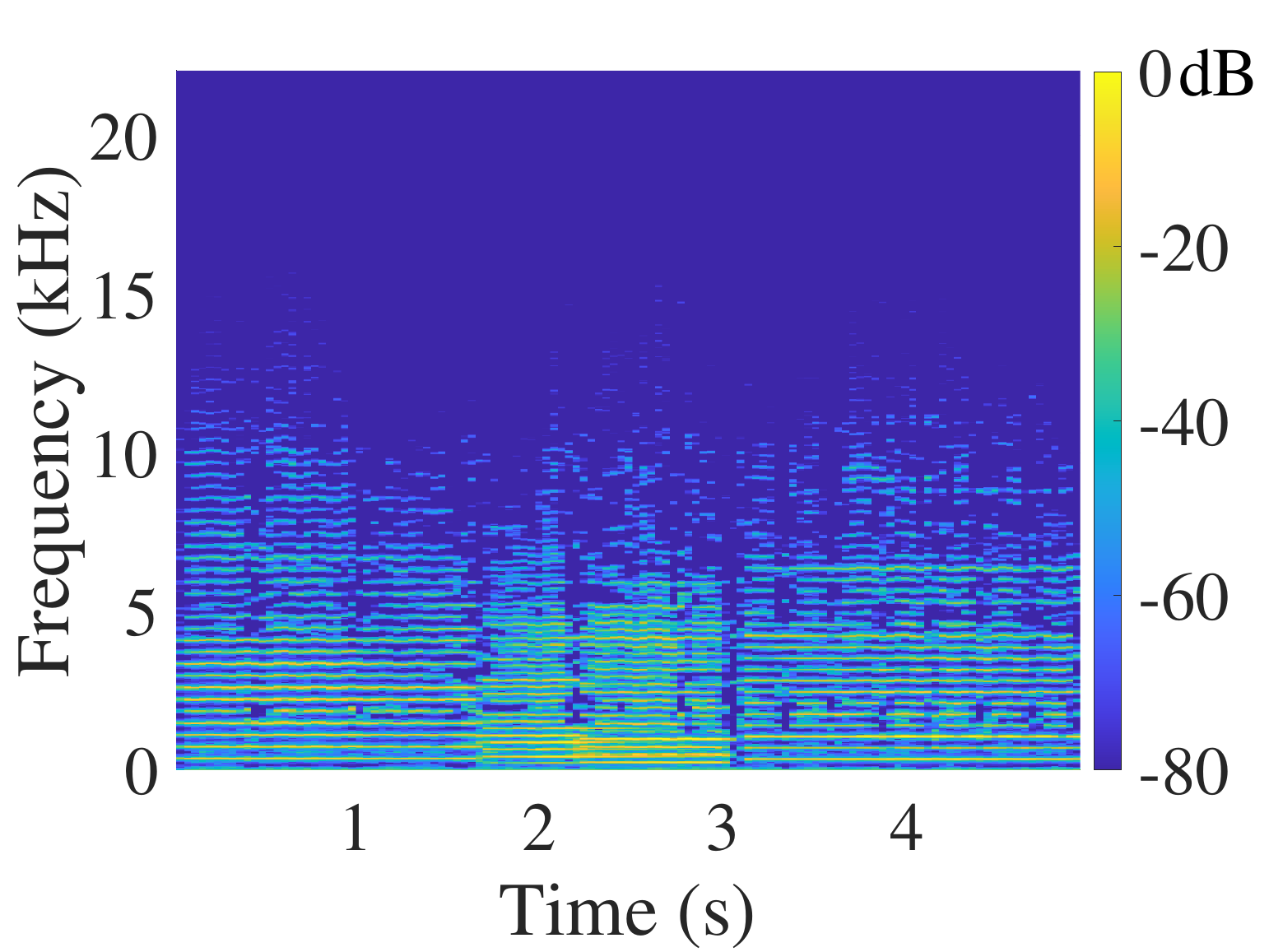}
\caption{Sines}
\end{subfigure}
\begin{subfigure}[t]{0.32\textwidth}
\centering
\includegraphics[width=\textwidth]{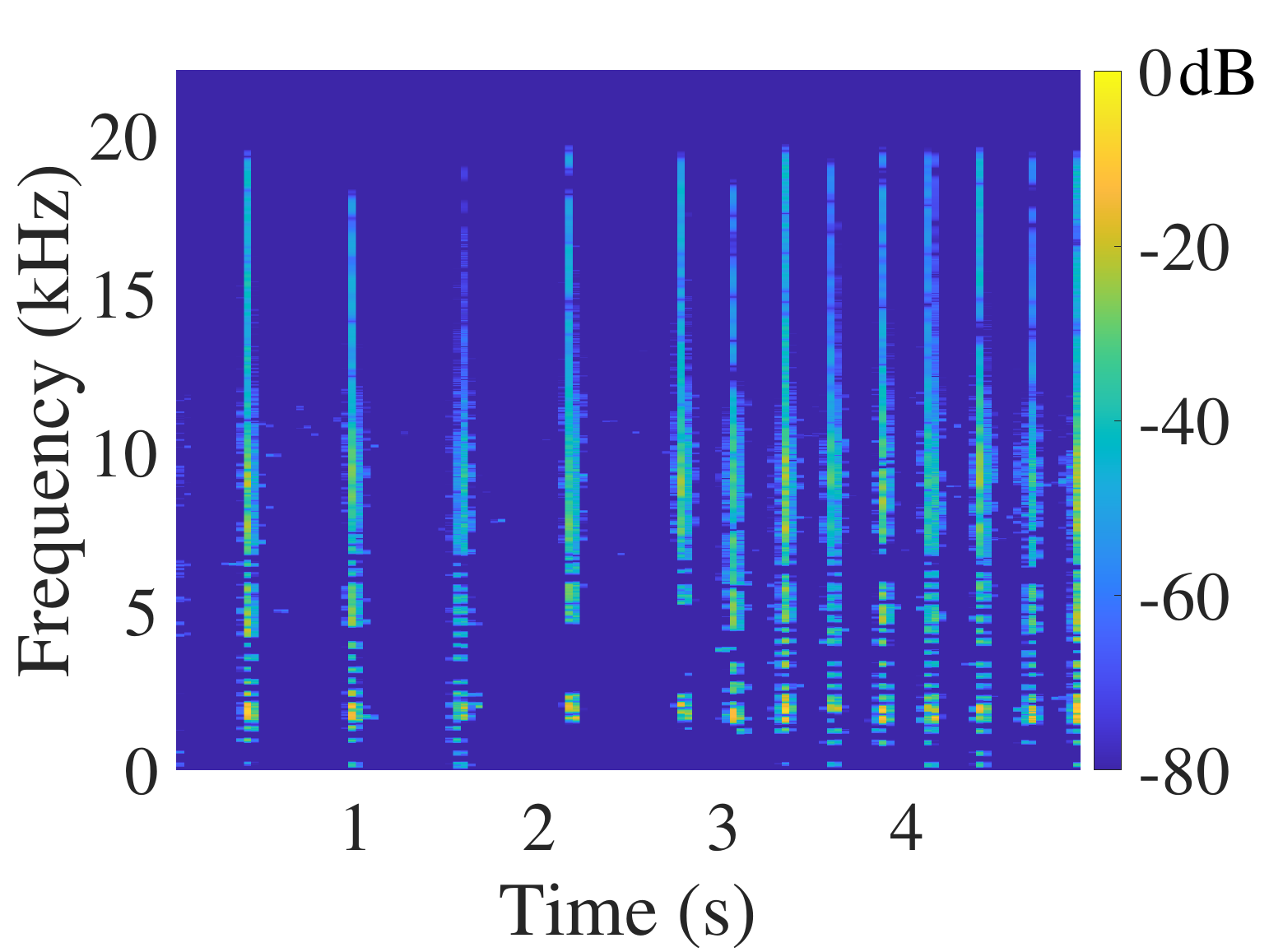}
\caption{Transients}
\end{subfigure}
\begin{subfigure}[t]{0.32\textwidth}
\centering
\includegraphics[width=\textwidth]{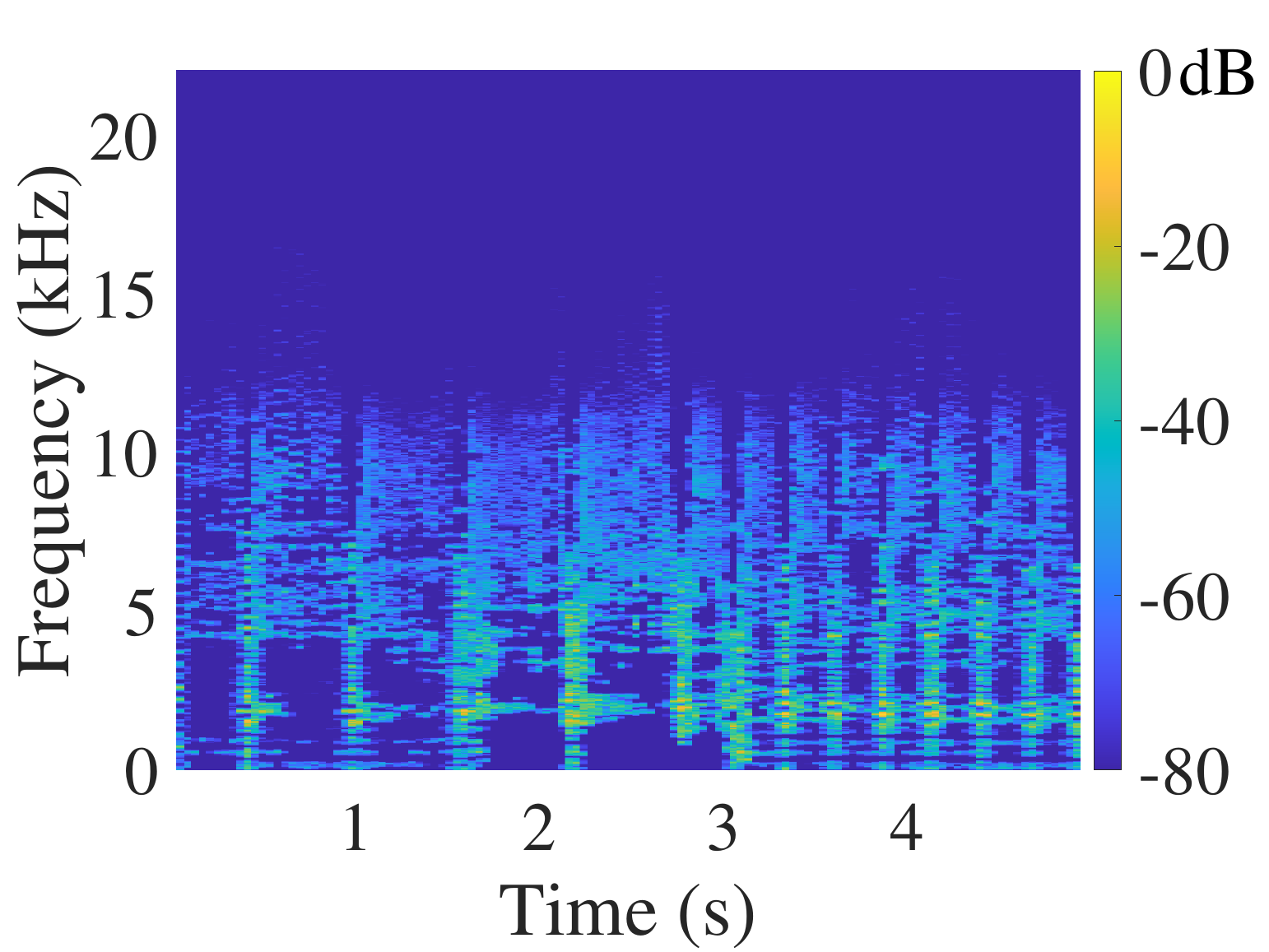}
\caption{Noise}
\end{subfigure}
\caption{\label{fig:STFTST}STN decomposition obtained using ST method, cf.~Fig.~\ref{fig:STFTHPR}.}
\end{figure*}

\noindent where $c$ is the anisotropy threshold and $r_\textrm{s}$ and $r_\textrm{t}$ are the frequency rate thresholds for the sinusoidal and the transient component, respectively. 
The noise mask is computed as described in Eq.~\eqref{eq:noise}, and the spectral components are then derived as in Eq.~\eqref{eq:masks}. 

Fig.~\ref{fig:STFTST} shows the separated STN components of the example audio signal using the ST method. Some differences can be observed in comparison to the separation results of the HPR method in Fig.~\ref{fig:STFTHPR}, such as some holes in the transient events at low and middle frequencies in Fig.~\ref{fig:STFTST}(b). 


\subsection{Fuzzy Separation}

Damskägg and Välimäki \cite{damskagg2017audio} introduced the concept of fuzzy classification of the spectral bins, which corresponds to a non--binary classification using continuous values between 0 and 1. This method was later extended by Moliner et al.~\cite{moliner2020virtual} to ensure perfect reconstruction, i.e. all masks summing up to unity. In \cite{moliner2020virtual}, a third membership function for noisiness $R_\textrm{n}$ is derived from Eqs.~\eqref{eq:tonalness} and \eqref{eq:transientness}:
\begin{equation} \label{eq:noisiness}
    R_\textrm{n}(m,k) = 1 - \sqrt{|R_\textrm{s}(m,k) -R_\textrm{t}(m,k)|}.
\end{equation}

\begin{figure}[t!]
\center
\includegraphics[width=0.86\columnwidth]{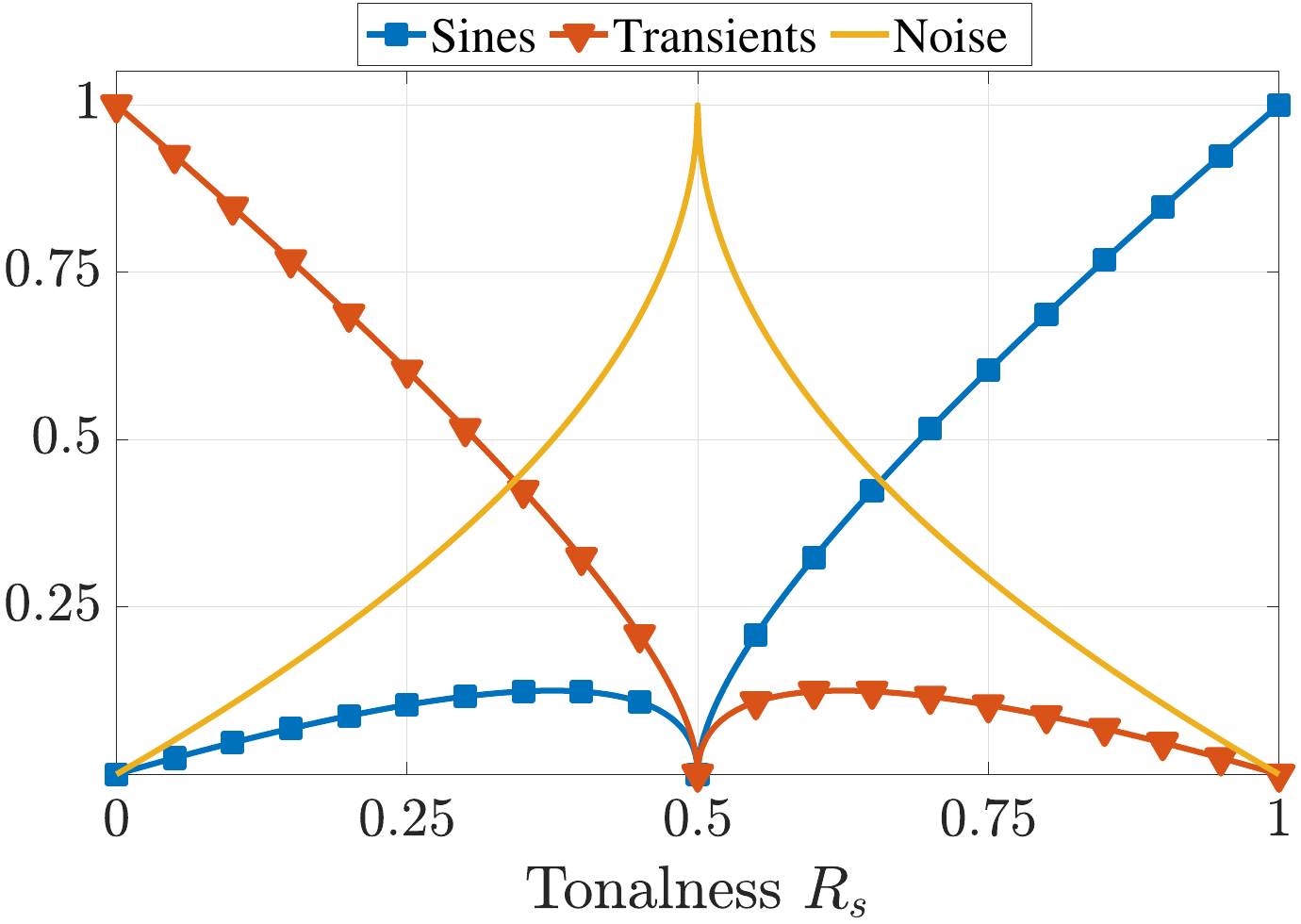}
\caption{\label{fig:fuzzy}{Transient, noise, and sinusoidal masks, as used in the FZ method, which ensures perfect reconstruction  \cite{moliner2020virtual}.}}
\end{figure}

\noindent The soft spectral masks are computed as
\begin{equation}
    S(m,k) =  R_\textrm{s}(m,k) - \frac{1}{2}\ R_\textrm{n}(m,k),
\end{equation}
\begin{equation}
    T(m,k) =  R_\textrm{t}(m,k) - \frac{1}{2}\ R_\textrm{n}(m,k), \\
\end{equation}
\noindent and
\begin{equation}
   N(m,k) = 1 - S(m,k) - T(m,k) = R_\textrm{n}(m,k).
\end{equation}
\noindent Their relationship is shown in Fig.~\ref{fig:fuzzy}. The spectral masks are once again imposed on $X(m,k)$ to obtain the spectral components using the Hadamard product, as in Eq.~\eqref{eq:masks}.

\begin{figure*}[t!]
\center
\begin{subfigure}[t]{0.32\textwidth}
\centering
\includegraphics[width=\textwidth]{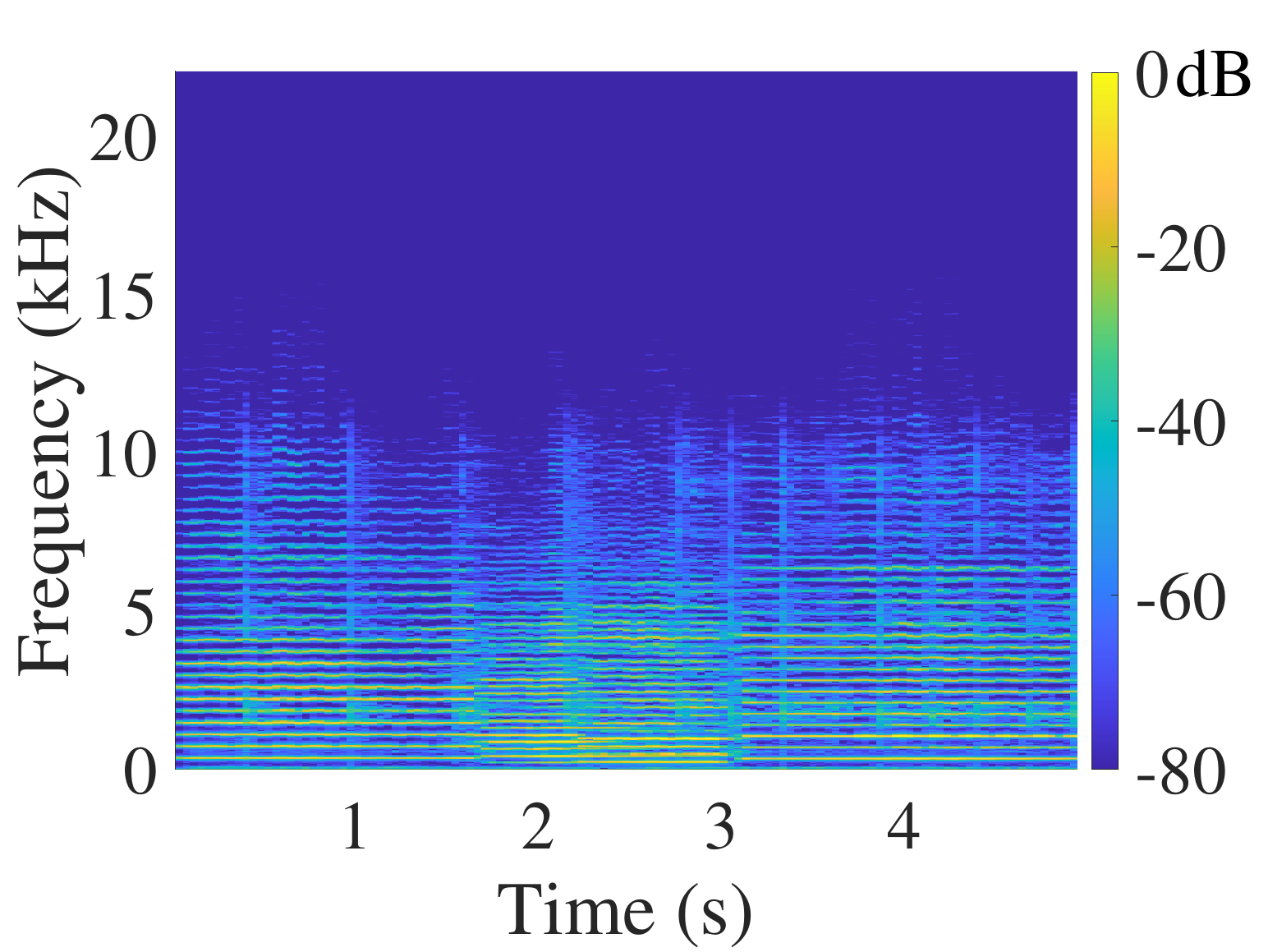}
\caption{Sines}
\end{subfigure}
\begin{subfigure}[t]{0.32\textwidth}
\centering
\includegraphics[width=\textwidth]{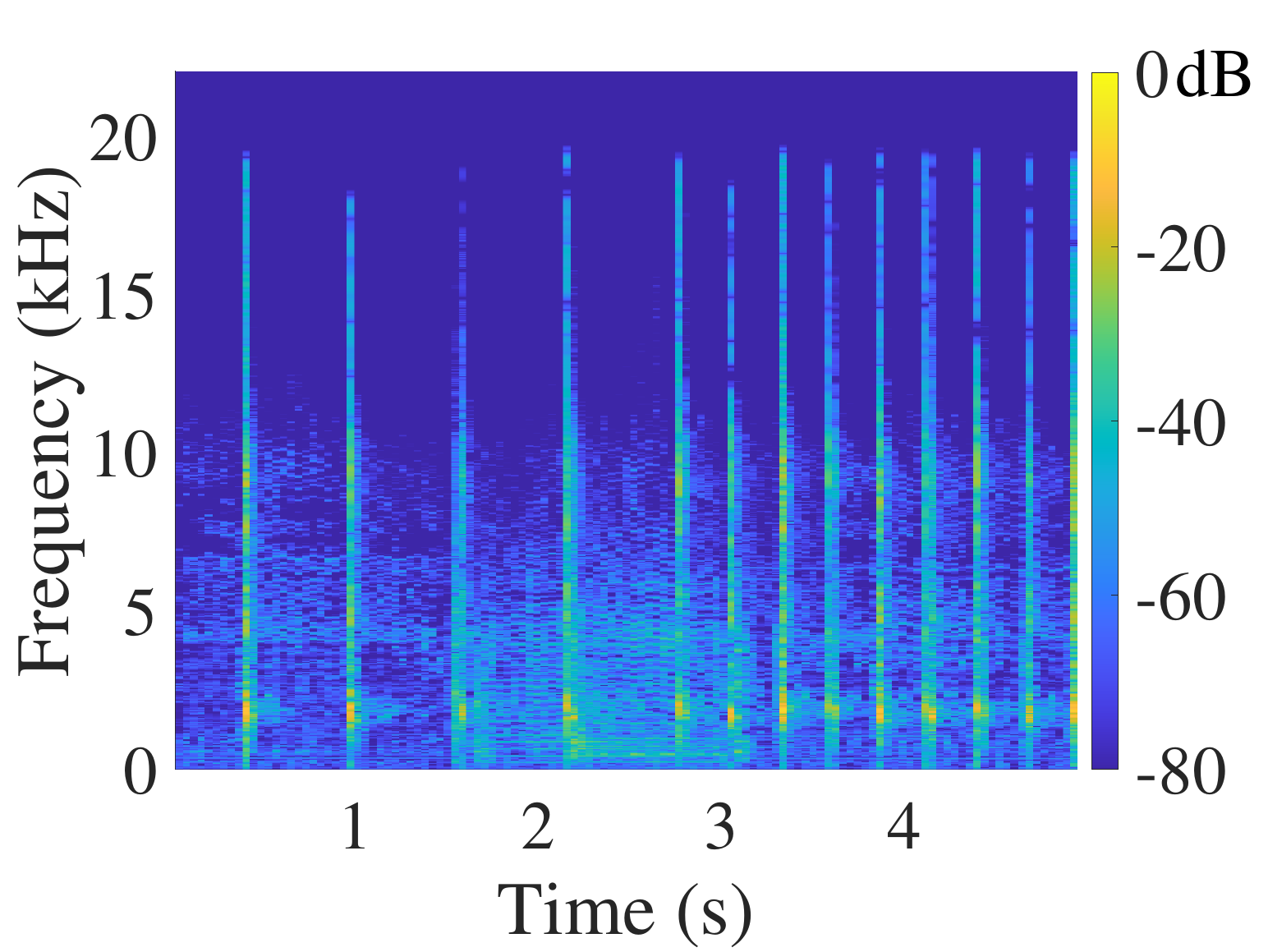}
\caption{Transients}
\end{subfigure}
\begin{subfigure}[t]{0.32\textwidth}
\centering
\includegraphics[width=\textwidth]{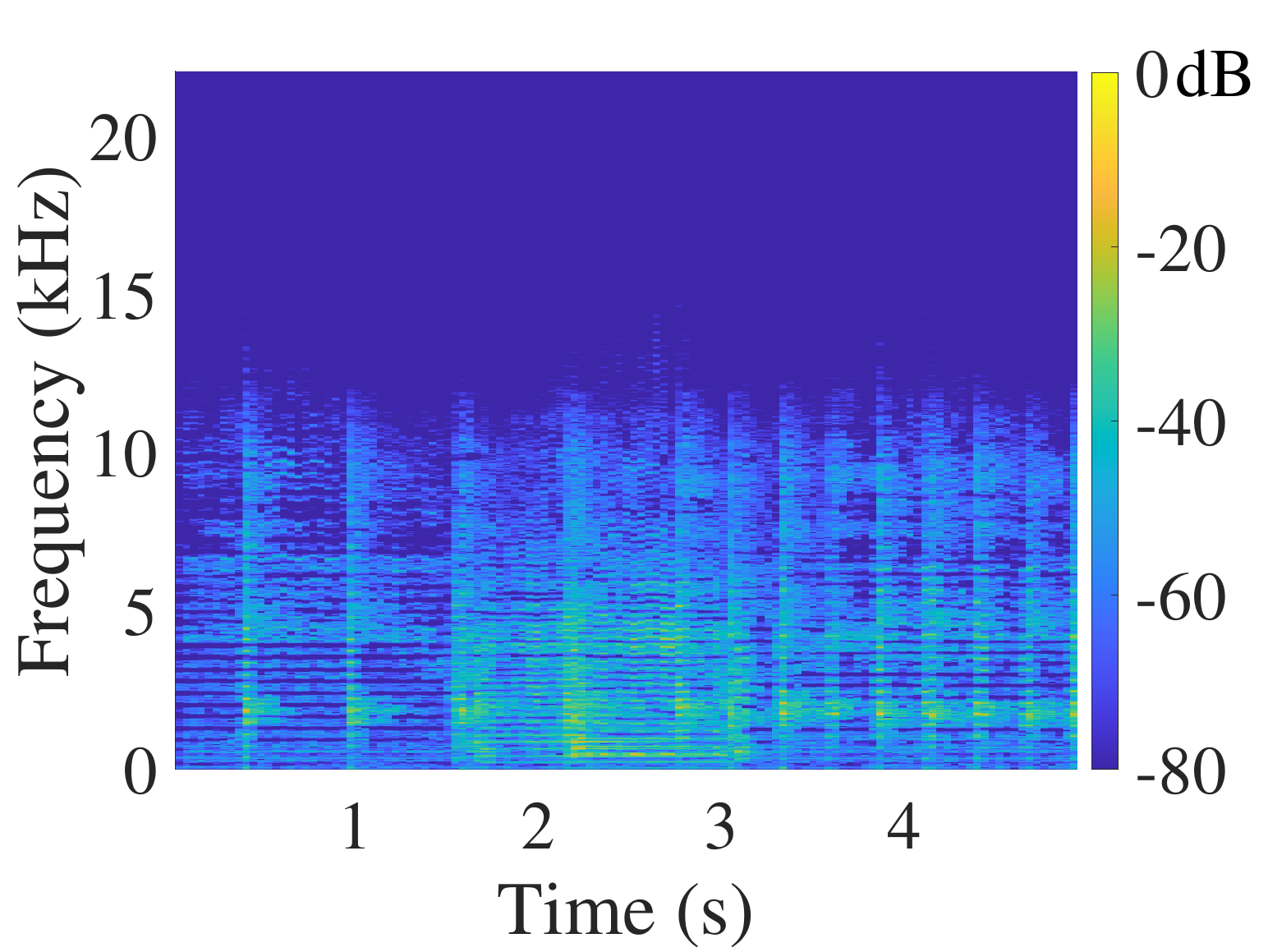}
\caption{Noise}
\end{subfigure}
\caption{\label{fig:STFTFZ}STN decomposition obtained using the FZ method, cf.~Figs.~\ref{fig:STFTHPR} and \ref{fig:STFTST}.}
\end{figure*}

Fig.~\ref{fig:STFTFZ} shows the separated STN components of the example signal using the fuzzy masks. The results are again slightly different from those obtained with the two previous techniques, presented in Figs.~\ref{fig:STFTHPR} and \ref{fig:STFTST}. One apparent feature is the leakage of energy from the other components to the transient component at frequencies below about 5\,kHz, shown in Fig.~\ref{fig:STFTFZ}(b). 

\section{PROPOSED METHOD}
\label{sec:method}

Fuzzy logic has proved useful in time-scale modification, where the separation of the STN components using a soft mask leads to the state-of-the-art performance \cite{damskagg2017audio,roberts2019time}. 
However, the mask for each component must be designed carefully to obtain the best performance. For example, the FZ separation proposed in \cite{moliner2020virtual} is suboptimal for this task, due to the leakage caused by the secondary lobes of the $S$ and $T$ masks and the peaky behavior of the $N$ mask, which can be seen in Fig.~\ref{fig:fuzzy}. Ideally, a good fuzzy masking approach guarantees perfect reconstruction, smooth transitions between classes, and a well defined dominant region per each class. In this section, a novel method comprising an extension to the HPR concept of clustered STN regions with soft masks resulting from fuzzy classification is proposed to fulfill this target. 

\subsection{Prototype Soft Masking}

\begin{figure}[t!]
\center
\includegraphics[width=0.86\columnwidth]{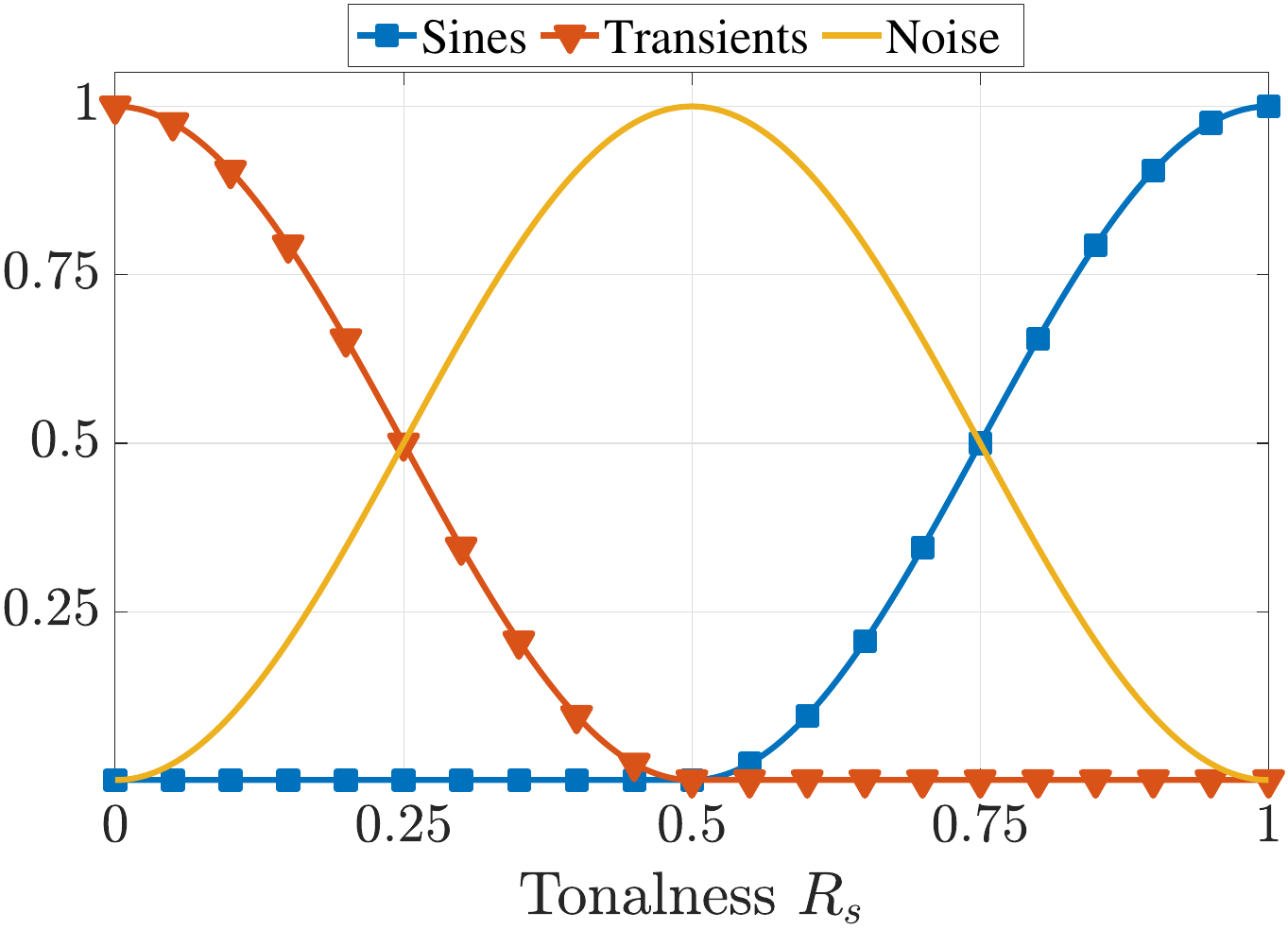}
\caption{\label{fig:hann}{Prototype soft masks for transients, noise, and sines, as computed from Eq.~\eqref{eq:hann}}.}
\end{figure}

A prototype function meeting all the aforementioned requirements is the raised--cosine function, also known as the Hann window:
\begin{equation}
    w(n) = \sin^2 (\pi\ n / L), \qquad 0\leq n<L.
\end{equation}
\noindent It is possible to take advantage of the symmetry of the raised--cosine function, using only its one wing (appropriately shifted) to describe the different transitions. The spectral masks for sines and transients can then be obtained as follows:
\begin{equation}\label{eq:hann}
\begin{aligned}
S(m,k) &= 
\begin{cases} 
\sin^2{[\pi (R_\textrm{s}(m,k) + \frac{1}{2})]}, & \mbox{if } R_\textrm{s}(m,k) \geq \frac{1}{2} \\
0, & \mbox{otherwise,}
\end{cases}\\
T(m,k) &= 
\begin{cases} 
\sin^2{[\pi (R_\textrm{s}(m,k) - \frac{1}{2})]}, & \mbox{if } R_\textrm{s}(m,k) \leq \frac{1}{2} \\
0, & \mbox{otherwise}
\end{cases}
\end{aligned}
\end{equation}

\noindent with $N(m,k)$ being computed according to Eq.~\eqref{eq:noise}. Their relationship is shown in Fig.~\ref{fig:hann}.
While the masks defined in Eq.~\eqref{eq:hann} already provide an audible improvement over the FZ masks, they are affected by a strong leakage of sines and transients into the noise component, suggesting that transitions between adjacent masks should be stricter. 

\subsection{Improving Noise Classification}
\label{sec:noiseclass}

Damsk\"agg and V\"alim\"aki \cite{damskagg2017audio} suggested that the tonalness distribution of pure noise, e.g. white or pink noise, can be used to verify the shape of the noise mask $N(m,k)$. In the following, a description of the noise distribution over the tonalness and, consequently,  sines-to-noise and transients-to-noise transitions is experimentally identified.

\begin{figure}[t!]
\center
\begin{subfigure}[t]{0.48\columnwidth}
\centering
\includegraphics[width=\columnwidth]{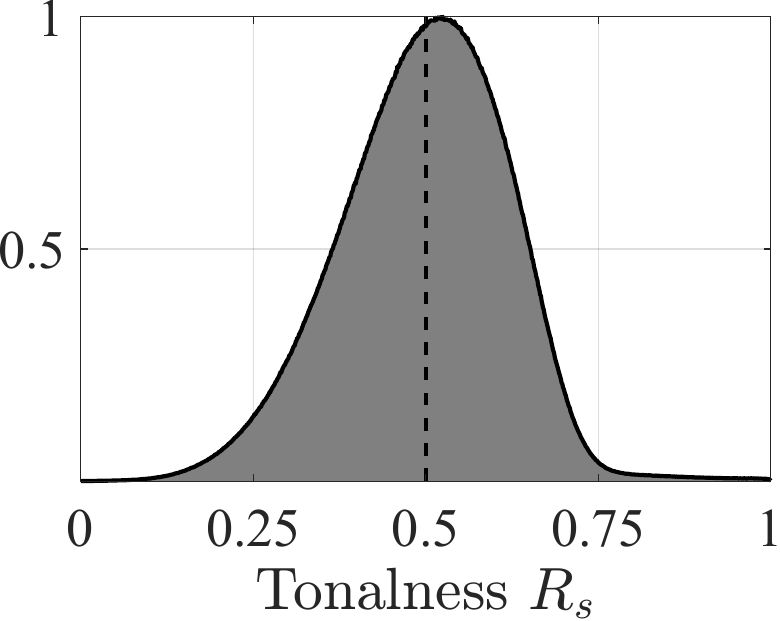}
\caption{\label{fig:NoiseLong}}
\end{subfigure}
\begin{subfigure}[t]{0.48\columnwidth}
\centering
\includegraphics[width=\columnwidth]{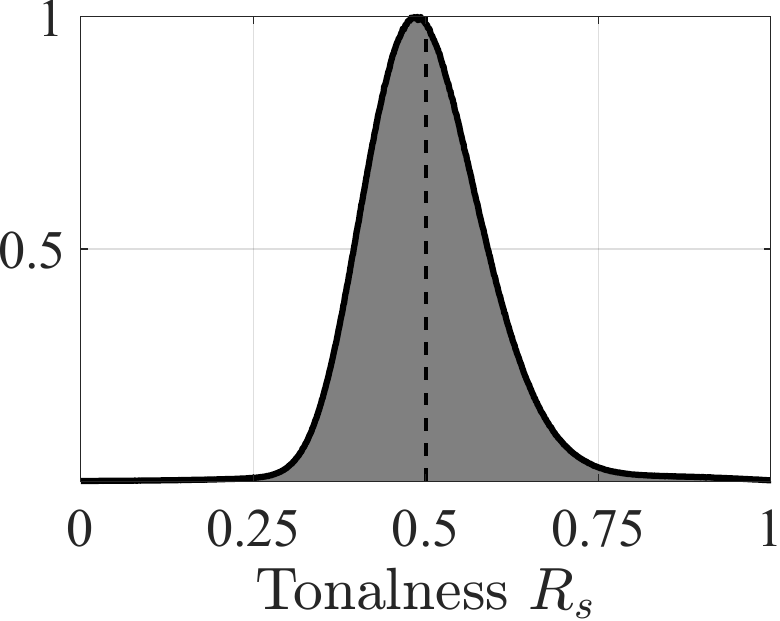}
\caption{\label{fig:NoiseShort}}
\end{subfigure}
\caption{Normalized tonalness distribution for median-filtered white noise with (a) long and (b) short analysis windows.}
\end{figure}

A set of 100 instances of random white noise were generated and their tonalness was computed, independently, with a long window ($185$\,ms, or $L = 8192$ samples at 44.1\,kHz) and a short window ($11$\,ms or $L = 512$ samples). Normalized histograms for tonalness values are shown, respectively, in Figs.~\ref{fig:NoiseLong} and \ref{fig:NoiseShort}.
A visual inspection indicates that the noise component remains relevant for a large range of tonalness values around 0.5 before quickly decaying in both directions. This suggests that mask transitions should be much steeper than in FZ (cf.~Fig.~\ref{fig:fuzzy}). The success of hard masking of HPR also indicates that a single-classification dominant region should be included in each mask.

It is also noted in Figs.~\ref{fig:NoiseLong} and \ref{fig:NoiseShort} that the shape of the tonalness distribution of noise is asymmetric for different window lengths. The peak value is not centered at $R_\textrm{s} = 0.5$, but shifts towards one side or another depending on $L$. The two wings of the distribution have different degrees of steepness: a sharper ``sinusoidal'' (right) side and a more relaxed ``transient'' (left) side can be identified for longer $L$  (Fig.~\ref{fig:NoiseLong}); the opposite behavior is exhibited for shorter $L$ (Fig.~\ref{fig:NoiseShort}). As for HPR, this could lead to a two--stage decomposition featuring masks with different transition regions.

\subsection{Enhanced Soft Masking with Fuzzy Logic}

\begin{figure}[t!]
\center
\includegraphics[width=0.86\columnwidth]{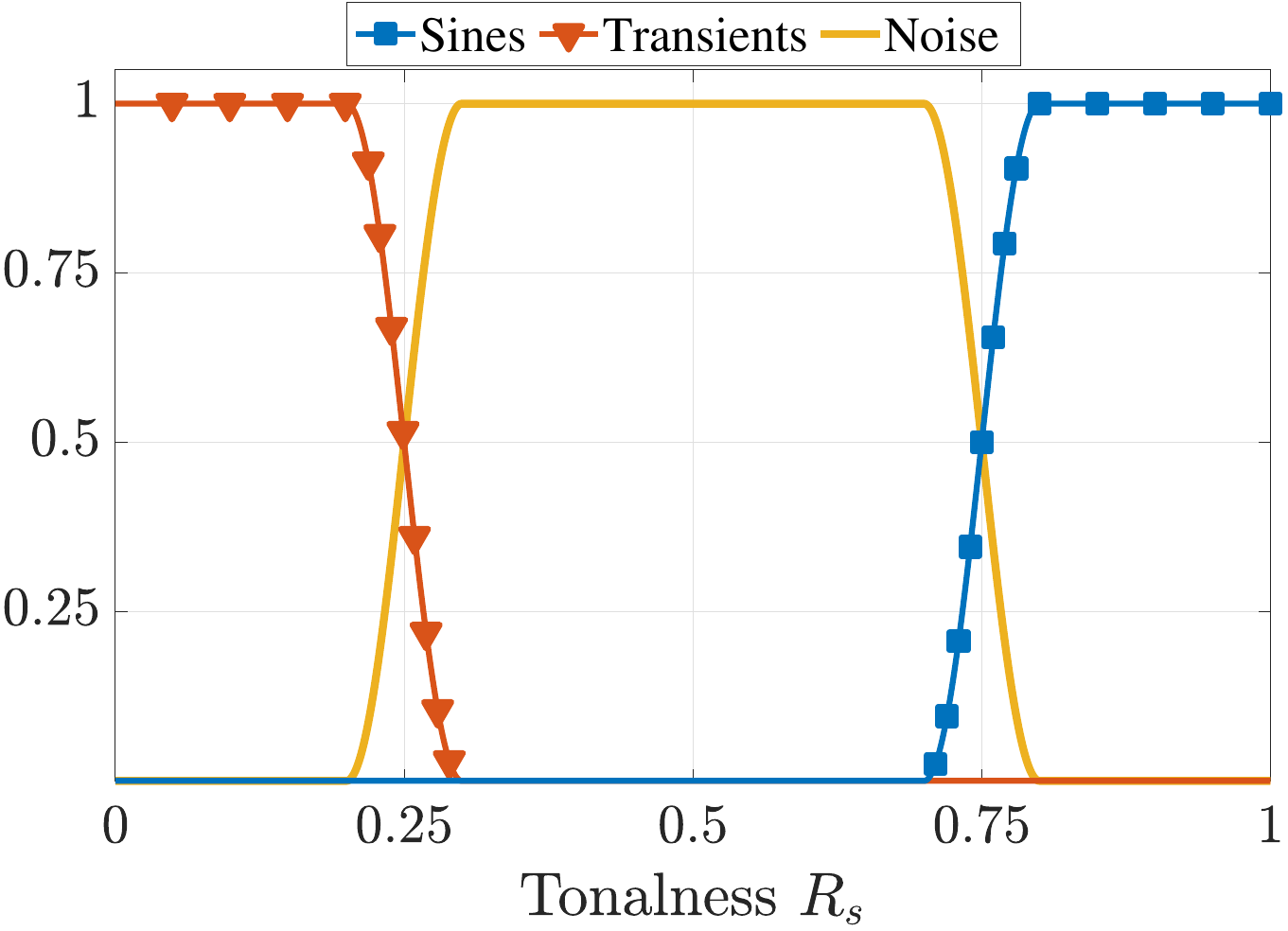}
\caption{\label{fig:prop}{Proposed enhanced transient, noise, and sines fuzzy masks for STN decomposition, $\beta_\textrm{U} = 0.8$, $\beta_\textrm{L} = 0.7$.}}
\end{figure}

Following the considerations discussed above, a new set of masks can be derived by altering Eq.~\eqref{eq:hann} to include dominant and cutoff regions for each mask while retaining the smoothness of the raised cosine function for the transition region. Parameters $\beta_\textrm{U}$ and $\beta_\textrm{L}$ are introduced to control the limits of the transition region and the bounds for, respectively, the dominant (upper) region  and the cutoff (lower) region. The enhanced fuzzy masks are obtained as follows:

\begin{equation}
S(m,k) = f\left(R_\textrm{s}(m,k)\right), \quad T(m,k) = f\left(R_\textrm{t}(m,k)\right)
\end{equation}
\noindent where
\begin{equation}
\begin{aligned}
f(&a) = 
\begin{cases} 
1, & \mbox{if } a \geq \beta_\textrm{U} \\
\sin^2{\Big( \dfrac{\pi}{2} \dfrac{a -\beta_\textrm{L}}{\beta_\textrm{U}-\beta_\textrm{L}}} \Big), & \mbox{if } \beta_\textrm{L} \leq a < \beta_\textrm{U} \\
0, & \mbox{otherwise}
\end{cases}
\end{aligned}
\end{equation}

\noindent and $N(m,k)$ is computed using Eq.~\eqref{eq:noise}. The relationship of the proposed soft masks is shown in Fig.~\ref{fig:prop}, when $\beta_\textrm{U} = 0.8$ and $\beta_\textrm{L}= 0.7$.

\begin{figure*}[t!]
\center
\includegraphics[width=\textwidth]{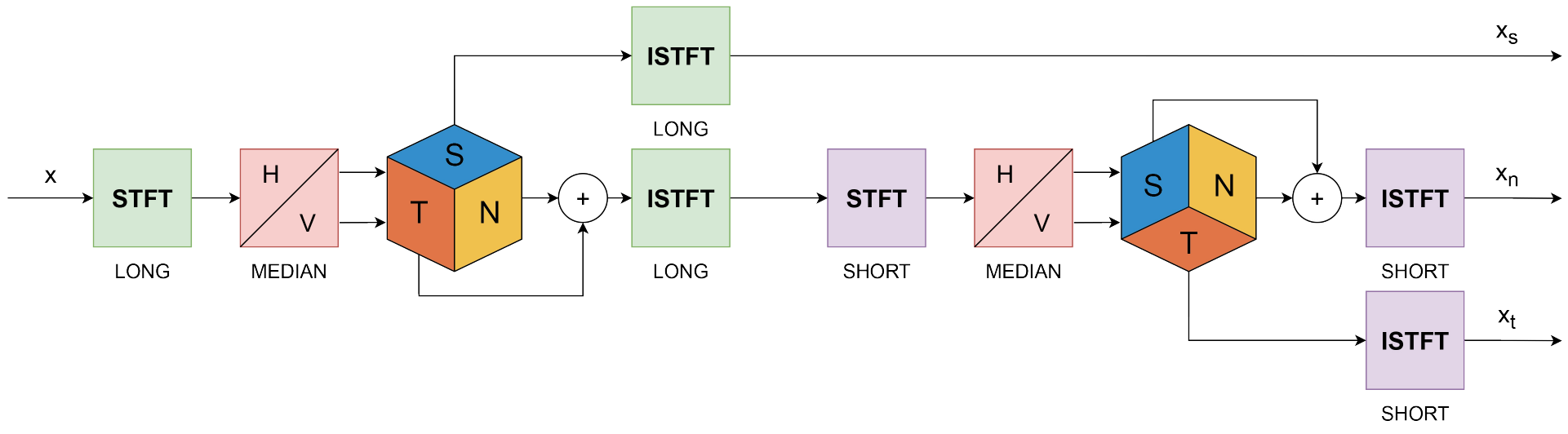}
\caption{\label{fig:diagram}{Block diagram of the proposed two--stage STN decomposition method.}}
\end{figure*}

It has been shown earlier that, with hard masks, the two--stage STN decomposition yields better results than a single--stage separation \cite{driedger2014extending,tachibana2013singing}. The same concept can be extended to fuzzy masks. Following Eqs.~\eqref{eq:firstround} and \eqref{eq:secondround}, two sets are obtained: $\{S_1,T_1,N_1,\beta_\textrm{U,1},\beta_\textrm{L,1}\}$ for sines extraction with a large $L$, and $\{S_2,T_2,N_2,\beta_\textrm{U,2},\beta_\textrm{L,2}\}$ for the residual transient and noise separation using a small $L$. The analysis process is summarized in Fig.~\ref{fig:diagram}.


\subsection{Choosing the Transition Area}

In order to find suitable values for the $\beta_\textrm{U}$ and $\beta_\textrm{L}$ parameters of the two decomposition stages, an optimization algorithm was run over the STN decomposition of a mixture of synthetic sounds, each belonging almost perfectly to a single class: a sum of sinusoids (for $S$), a short Gaussian monopulse\footnote{$x_{\text{GP}}(t)=\sqrt{e} 2\pi f_c t e^{-2 (\pi f_\textrm{c} t)^2}$, where $f_\textrm{c}$ is the center frequency.} (for $T$), and a white noise sequence (for $N$). As the original sources are known, the decomposition error can be evaluated for each class. The goal is not to find a single optimal pair of values for the interval, as it is known that the quality of STN decomposition greatly varies with different audio inputs \cite{driedger2014extending, damskagg2017audio}. Instead, a range of tonalness values for each bound yielding a small--enough decomposition error can be identified; the following analysis was conducted in order to find a pair of quasi--optimal values that suits multiple audio inputs.

The genetic algorithm \cite{mirjalili2019genetic} was chosen for the optimization process, which is divided in two stages. The first optimization run narrows down a set of paired bounds $B_1 =  \{ (\beta_\textrm{U,1}, \beta_\textrm{L,1})_i\} $ over the $S_1$ mask by minimizing the decomposition error over the sinusoidal part. Following that, different optimizations can be run by fixing a single pair $ \{\beta_\textrm{U,1}, \beta_\textrm{L,1}\}$ from $B_1$ and finding its optimal pair $\{\beta_\textrm{U,2}, \beta_\textrm{L,2}\}$ over the $T_2$ mask. Finally, an audible comparison over multiple separations using the obtained sets determines the final quasi--optimal set $\Bar{B} = \{(\beta_\textrm{U,1}, \beta_\textrm{L,1}), (\beta_\textrm{U,2}, \beta_\textrm{L,2})\}$, which ensures that the decomposition quality remains similar for different audio inputs. The results of this quasi--optimal choosing process are reported in Table \ref{tab:optimal}. 

\addtolength{\tabcolsep}{5pt}
\begin{table}[t!]
\centering
\caption{Transition area bounds for each decomposition stage used in this study.}
\begin{tabular}{llll}
\hline
\hline
Stage & Decomposition & $\beta_\textrm{U}$ & $\beta_\textrm{L}$ \\ \hline
1  & Sines v Residual    & 0.8                & 0.7                \\ 
2 & Transients v Noise    & 0.85               & 0.75                                    \\ \hline
\end{tabular}
\label{tab:optimal}
\end{table}
\addtolength{\tabcolsep}{-5pt}

\begin{figure*}[t!]
\center
\begin{subfigure}[t]{0.32\textwidth}
\centering
\includegraphics[width=\textwidth]{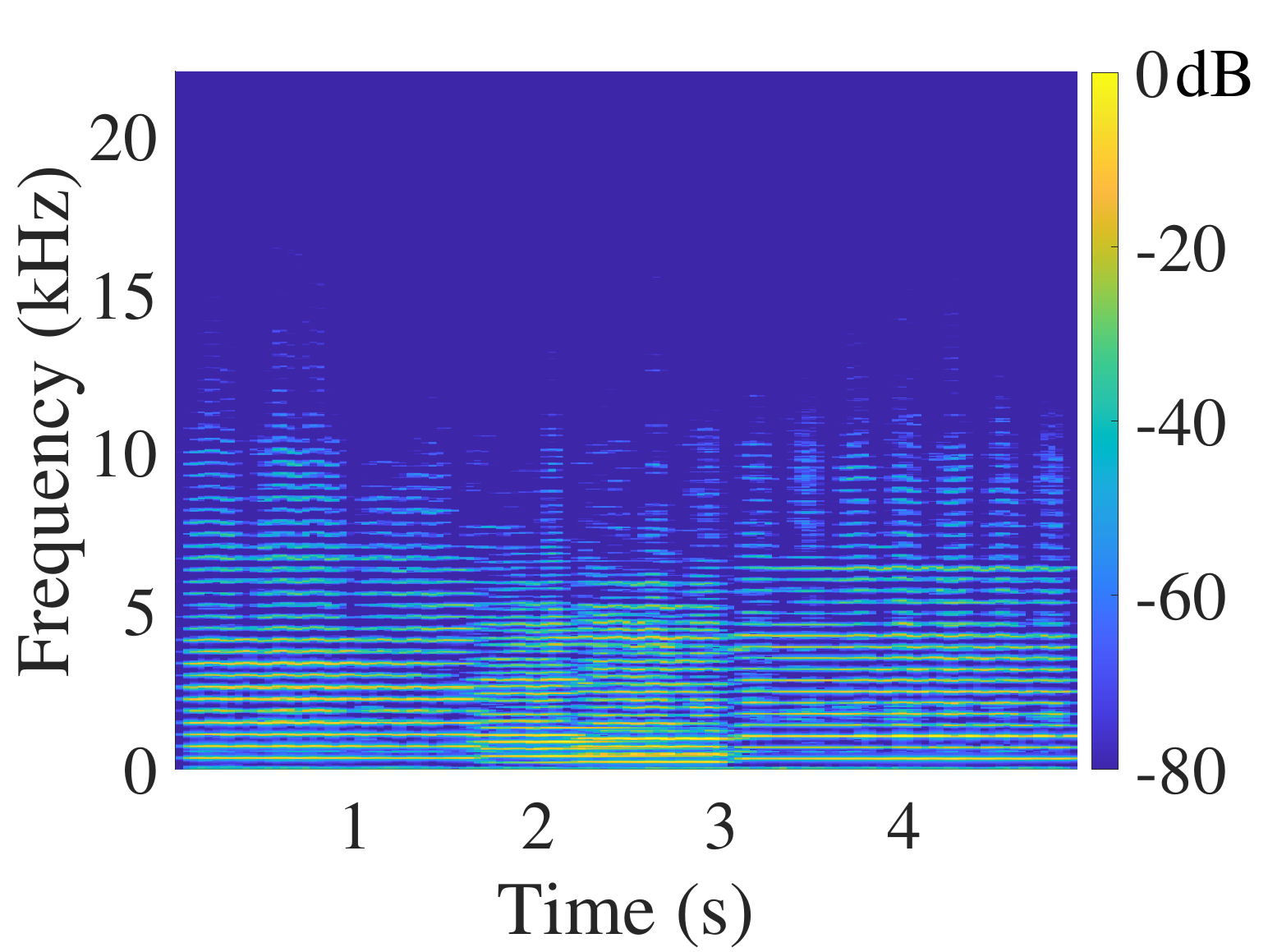}
\caption{Sines}
\end{subfigure}
\begin{subfigure}[t]{0.32\textwidth}
\centering
\includegraphics[width=\textwidth]{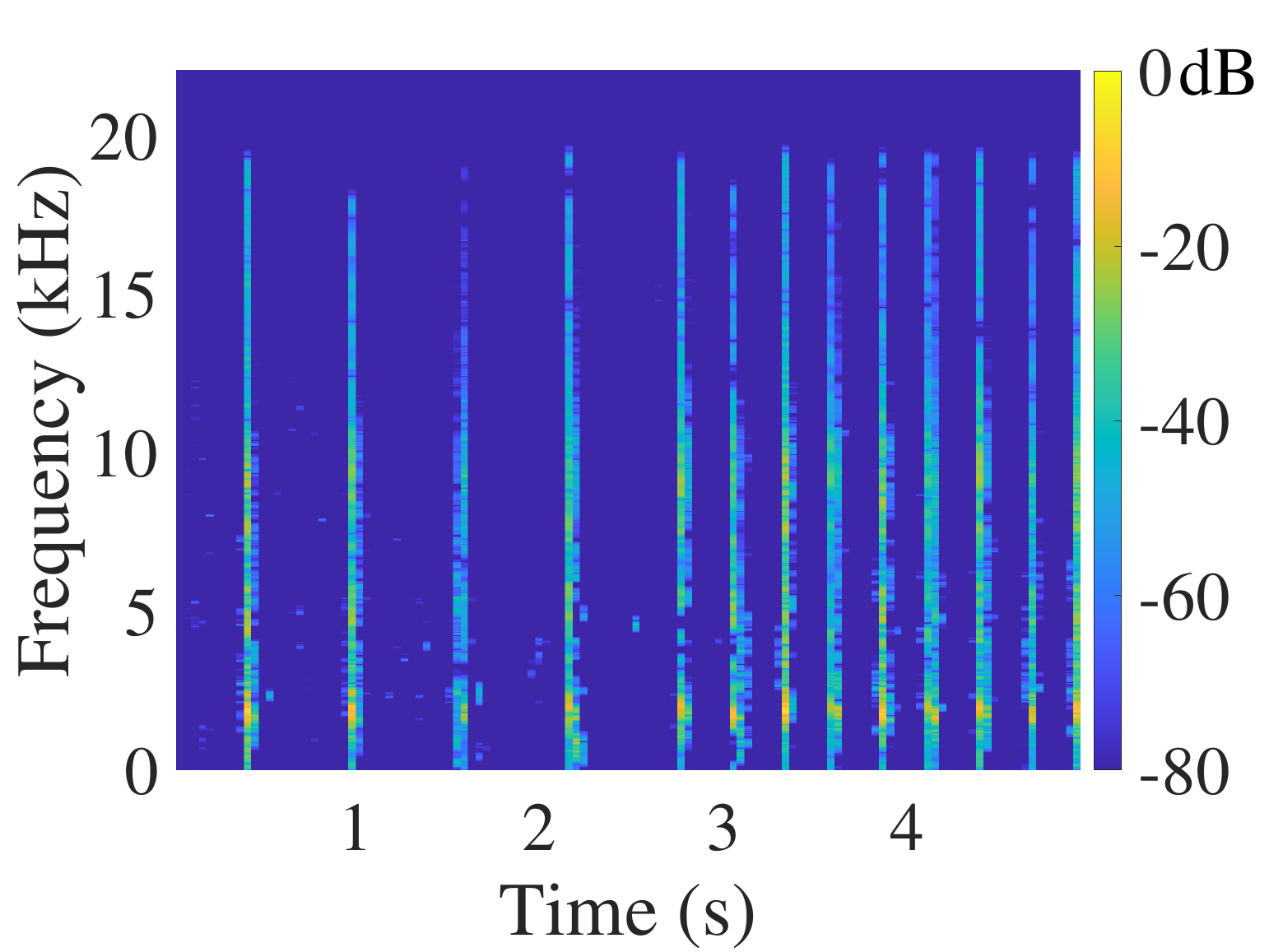}
\caption{Transients}
\end{subfigure}
\begin{subfigure}[t]{0.32\textwidth}
\centering
\includegraphics[width=\textwidth]{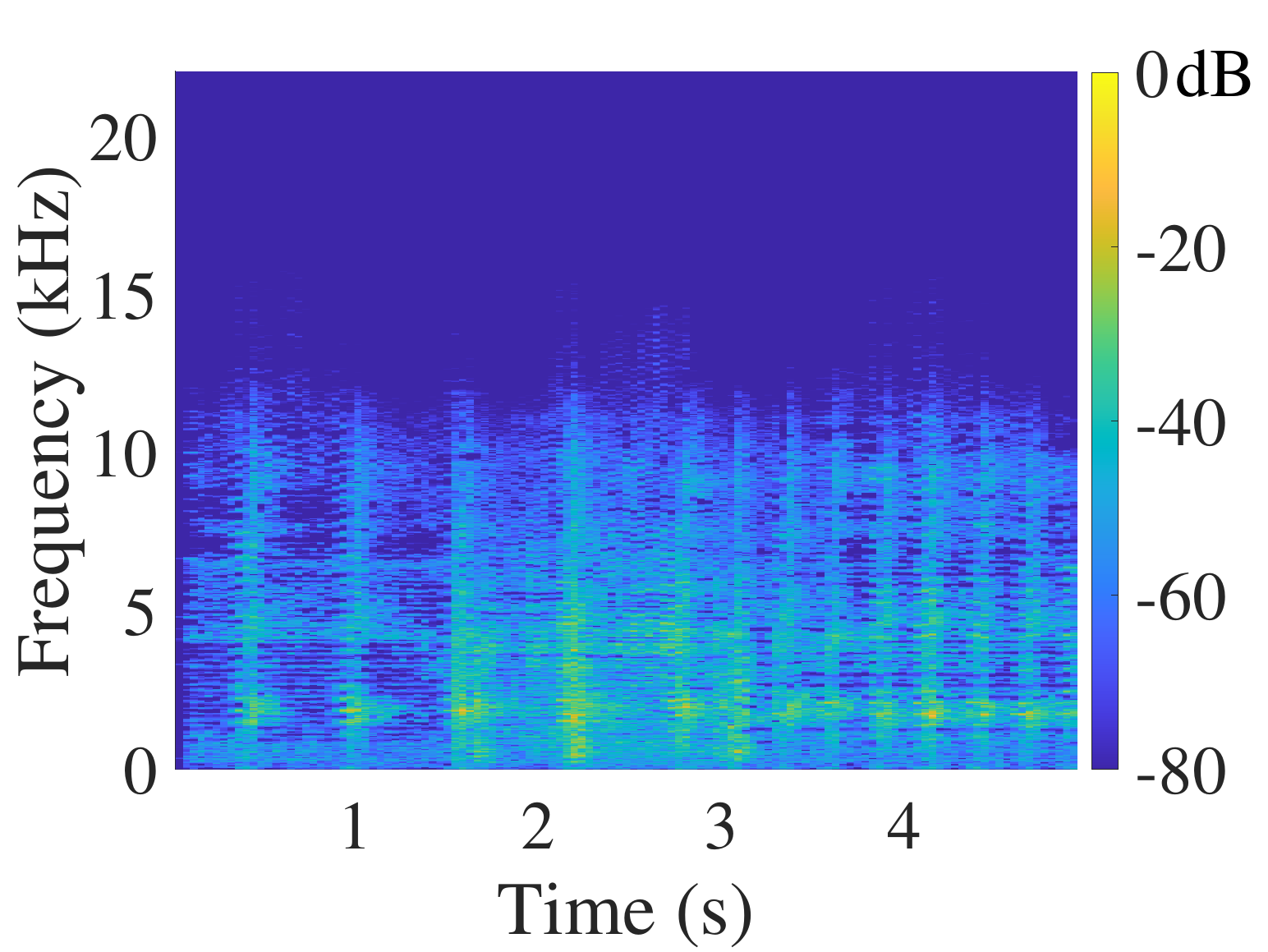}
\caption{Noise}
\end{subfigure}
\caption{\label{fig:STFTPROP}STN decomposition of the castanets and violin using the proposed method, cf.~Figs.~\ref{fig:STFTHPR}, \ref{fig:STFTST}, and \ref{fig:STFTFZ}.}
\end{figure*}

Fig.~\ref{fig:STFTPROP} shows the separated STN components of the example audio signal using the proposed method with the chosen set $\Bar{B}$. The separation results differ somewhat from the previous separation examples. In particular, the transients in Fig.~\ref{fig:STFTPROP}(b) are unbroken and the pauses between them are practically free from leakage from the other components, as desired. 

\section{EVALUATION}
\label{sec:evaluation}

The audio quality of STN decomposition is typically degraded by inter-component leakage, loss of tonality, loss of presence, or other artifacts, e.g.~musical noise \cite{fierro2021sitrano}. In previous works, the separation quality was evaluated by means of audio blind source separation performance assessment metrics, such as SDR (Signal-to-Distortion Ratio), SIR (Signal-to-Interference Ratio), and SAR (Signal-to-Artifacts Ratio) \cite{driedger2014extending, fug2016harmonic,vincent2006performance}. However, those metrics require a mixture of three independent sources, one per class, that is subsequently decomposed again for the separation quality assessment. This prevents non-synthetic audio inputs, e.g. music or speech, from being tested: unless dealing with perfectly tonal, impulsive, or noisy sources, the input sounds themselves are composed of a mixture of unknown sinusoidal, transient, and noise parts. 


Informal subjective listening tests on STN separation have previously been conducted by asking the participants whether the sinusoidal and transient components extracted by the method under test met the expectation of representing the sines and transients of the audio reference \cite{driedger2014extending}.
The STN decomposition algorithm proposed in this work is evaluated against other techniques by extending the same idea to a formal blind listening test, involving experienced listeners as participants and asking them to rate the quality of the sines and transients extraction for different STN methods.



\subsection{Listening Test Design}
\label{sec:testdesign}

A formal blind listening test was conducted on a selection of 19 experienced listeners, 17 of which reported previous experience in test design. No participant reported any hearing impairments or relevant medical conditions. 
The test software was run on a machine running MacOS 10.14.6, using a single pair of Sennheiser HD 650 headphones, inside a sound--proof listening booth at the Aalto Acoustics Lab, Espoo, Finland. 

A set of nine audio samples of short duration (4 to 6\,s) was selected, consisting of two synthetic sounds and seven musical excerpts from various genres, featuring different spectral contents. The test samples are listed in Table \ref{tab:samples}.

\begin{table}[t!]
\centering
\caption{Audio samples used in the listening test.}
\resizebox{\columnwidth}{!}{%
\begin{tabular}{ll}
\hline
\hline
Name & Description \\ \hline
Synth                & Synthetic mix of tones, pulses and white noise \\ 
CastViol             & Solo violin and castanets, from \cite{driedger2014tsm}                                      \\ 
ICanSee              & Excerpt from \textit{I Can See Clearly}, by \textit{Holly Cole Trio}                            \\ 
Eddie                & Excerpt from \textit{Early in the Morning}, by \textit{E. Rabbit}                                \\ 
Jazz                 & Mix of trumpet, piano, bass, and drums, from \cite{driedger2014tsm}                                \\ 
Vocals               & Excerpt from \textit{Tom’s Diner}, by \textit{Suzanne Vega}                                         \\ 
Vibrato              & Synthetic mix of vibrato, pulses, and pink noise        \\
Billie               & Intro of \textit{Billie Jean}, by \textit{Michael Jackson}                    \\ 
Drum                 & Solo performed on a drum set, from \cite{driedger2014tsm}                                   \\ \hline
\end{tabular}}
\label{tab:samples}
\end{table}


\begin{figure*}[t!]
\center
\begin{subfigure}[t]{\textwidth}
\includegraphics[width=\textwidth]{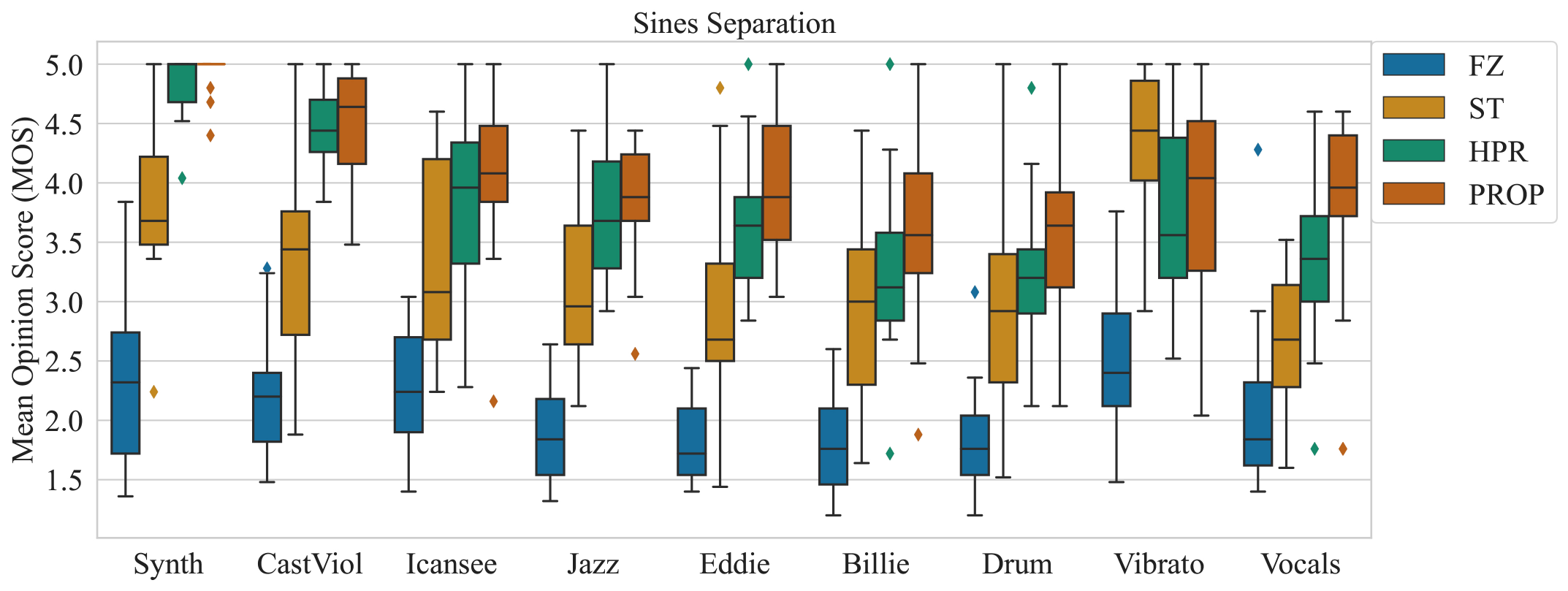}

\caption{\label{fig:BoxSines}}
\vspace*{3mm}
\end{subfigure}

\begin{subfigure}[t]{\textwidth}


\includegraphics[width=\textwidth]{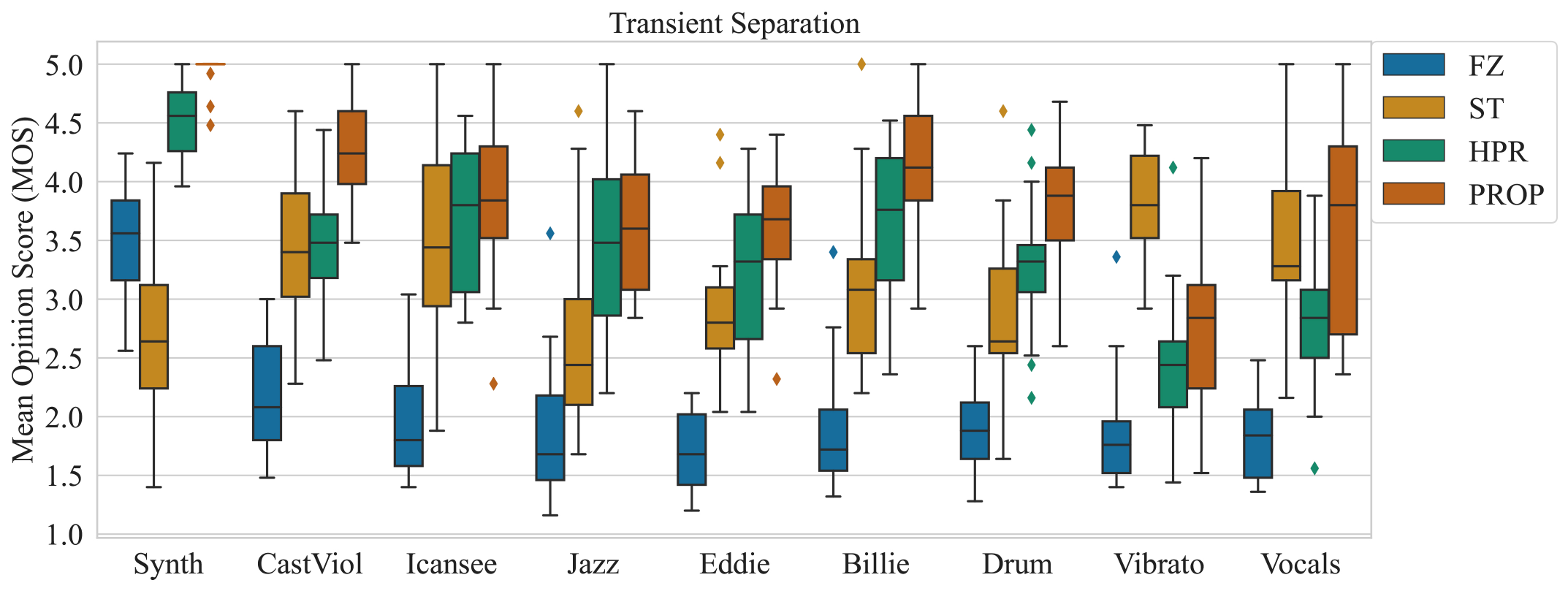}
\caption{\label{fig:BoxTransient}}
\end{subfigure}
\caption{\label{fig:eval}{Mean opinion scores and confidence intervals for (a) sines and (b) transients separation of nine audio samples, showing that the proposed method is the winner or among the best methods in almost all cases.}}
\end{figure*}

In each trial of the test, subjects were presented with one of the audio excerpts, referred to as \textit{reference}, and were asked to blindly rate the quality of the extraction of the sound component under test (sines or transients, respectively) from such a reference, for four different STN decomposition methods: HPR, ST, FZ and the proposed one (PROP). The original reference was also included among the samples under test, to provide a lower bound. Subjects were asked to rate each sample on a scale from 0 to 100, with the specific request of assigning 0 to the sample identified as the reference. 
As there was no ground truth separation for any of the synthetic sounds, it is anticipated that none of the samples was perceived to be an ideal decomposition: hence, there was no obligation to rate any of the samples as 100 (full scale). 

The test was divided in two parts consisting of nine trials each, the first focusing on sines extraction and the second on transients extraction, for a total of 18 trials and 90 audio samples under investigation.
Listeners were allowed a short training session before starting the actual test to get acquainted with the interface, the keyboard shortcuts, and the task itself. The results of the training session were not included in the statistical analysis. Prior to the test, familiarity of the subjects with the concepts of sound separation and sines, transients, and noise was asserted.
The experiment was conducted over WebMushra \cite{schoeffler2018webmushra}, although the designed test did not follow the MUSHRA recommendation. The processed audio excerpts and the test software are available at the companion webpage \cite{webpage}.

\subsection{Results}

\begin{figure*}[t!]
	\centering
	\begin{subfigure}[t]{0.15\textwidth}
		\centering
		\includegraphics[width=\hsize]{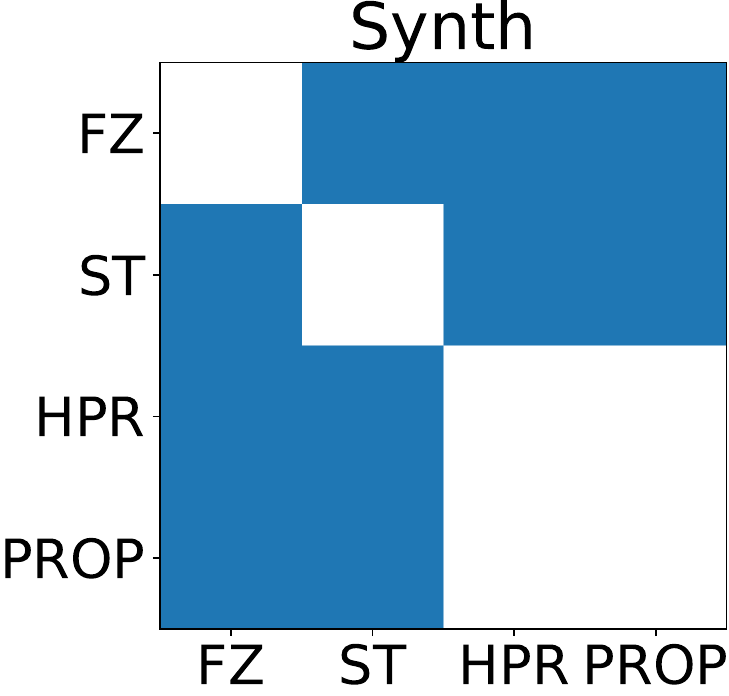}
		\caption{}
		\vspace*{4mm}
	\end{subfigure}
	\hspace{0.03\textwidth}
		\begin{subfigure}[t]{0.15\textwidth}
		\centering
		\includegraphics[width=\hsize]{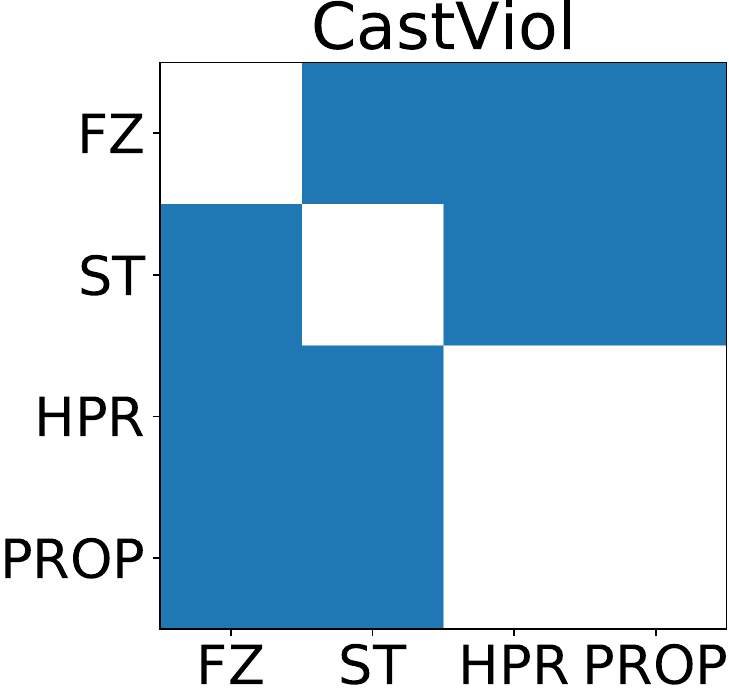}
		\caption{}
		\vspace*{4mm}
	\end{subfigure}
	\hspace{0.03\textwidth}
		\begin{subfigure}[t]{0.15\textwidth}
		\centering
		\includegraphics[width=\hsize]{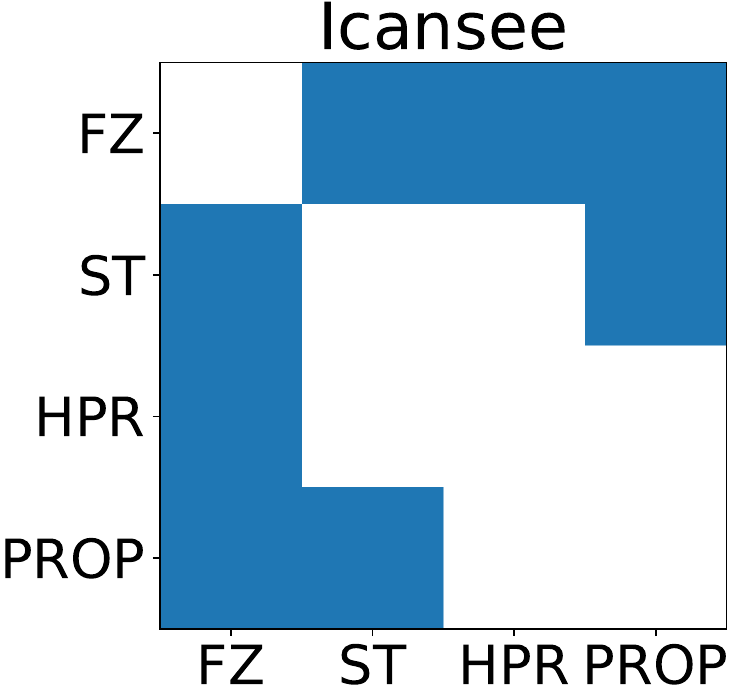}
		\caption{}
		\vspace*{4mm}
	\end{subfigure}
		\hspace{0.03\textwidth}
		\begin{subfigure}[t]{0.15\textwidth}
		\centering
		\includegraphics[width=\hsize]{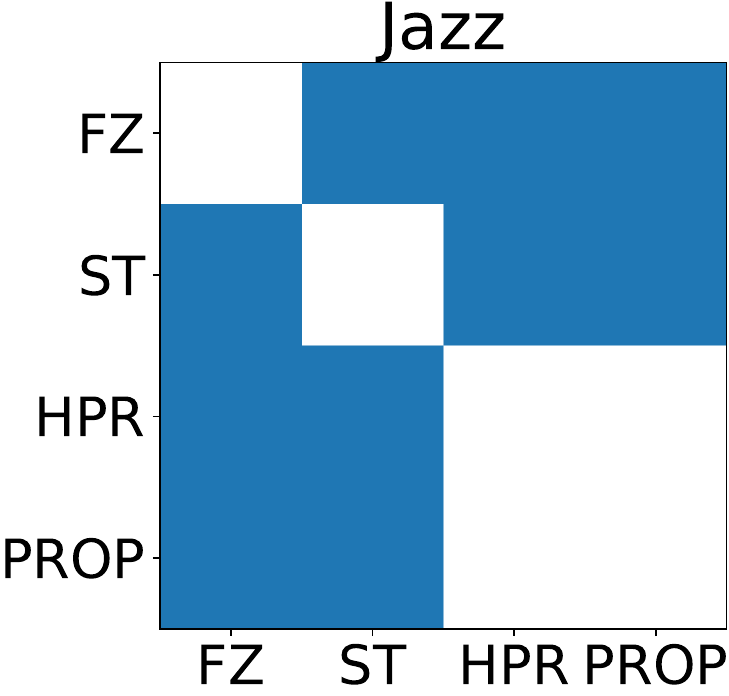}
		\caption{}
		\vspace*{4mm}
	\end{subfigure}
	\hspace{0.03\textwidth}
	\begin{subfigure}[t]{0.15\textwidth}
		\centering
		\includegraphics[width=\hsize]{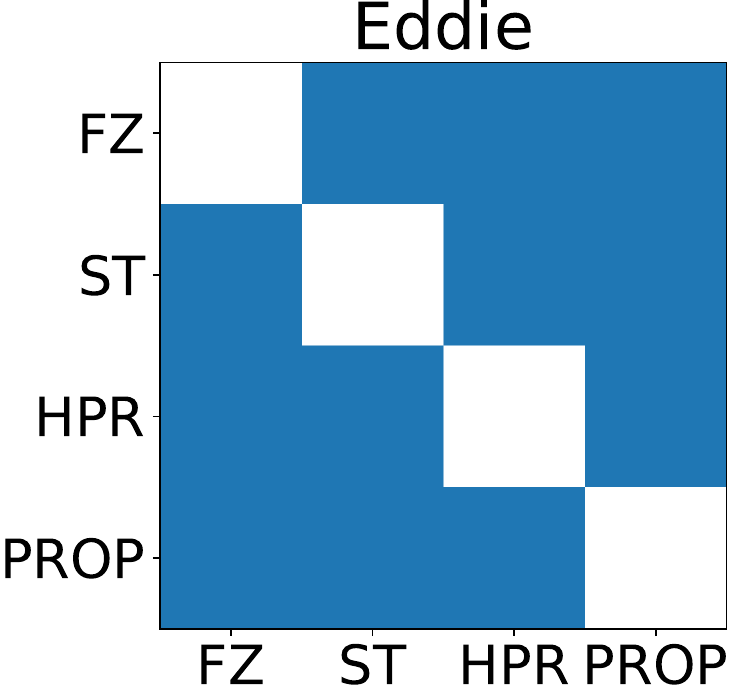}
		\caption{}
		\vspace*{4mm}
	\end{subfigure}
	\hspace{0.03\textwidth}
		\begin{subfigure}[t]{0.15\textwidth}
		\centering
		\includegraphics[width=\hsize]{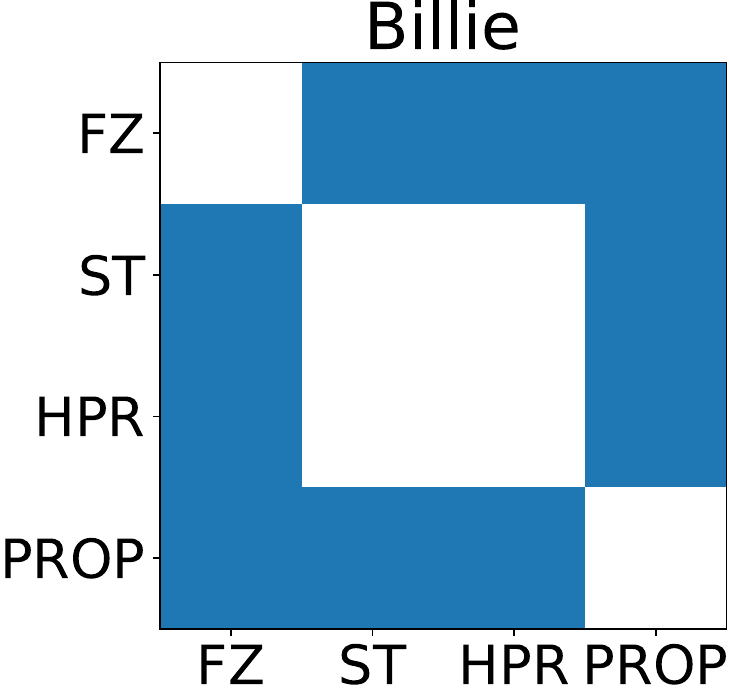}
		\caption{}
	\end{subfigure}
	\hspace{0.045\textwidth}
		\begin{subfigure}[t]{0.15\textwidth}
		\centering
		\includegraphics[width=\hsize]{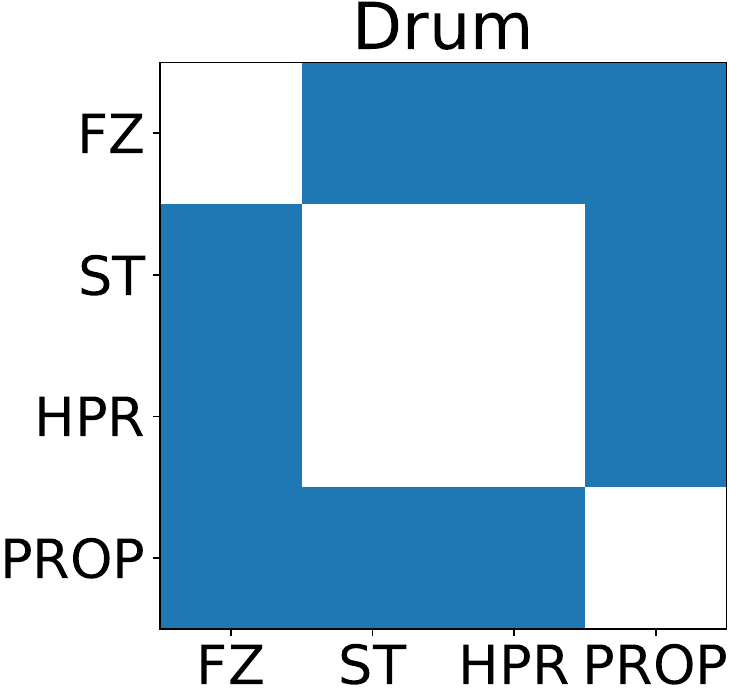}
		\caption{}
	\end{subfigure}
	\hspace{0.03\textwidth}
	\begin{subfigure}[t]{0.15\textwidth}
		\centering
		\includegraphics[width=\hsize]{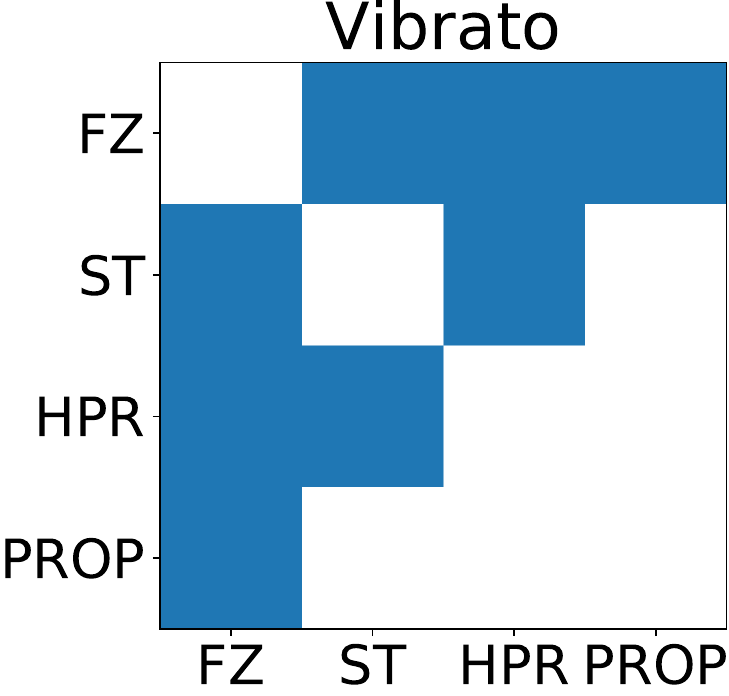}
		\caption{}
	\end{subfigure}
	\hspace{0.03\textwidth}	
	\begin{subfigure}[t]{0.15\textwidth}
		\centering
		\includegraphics[width=\hsize]{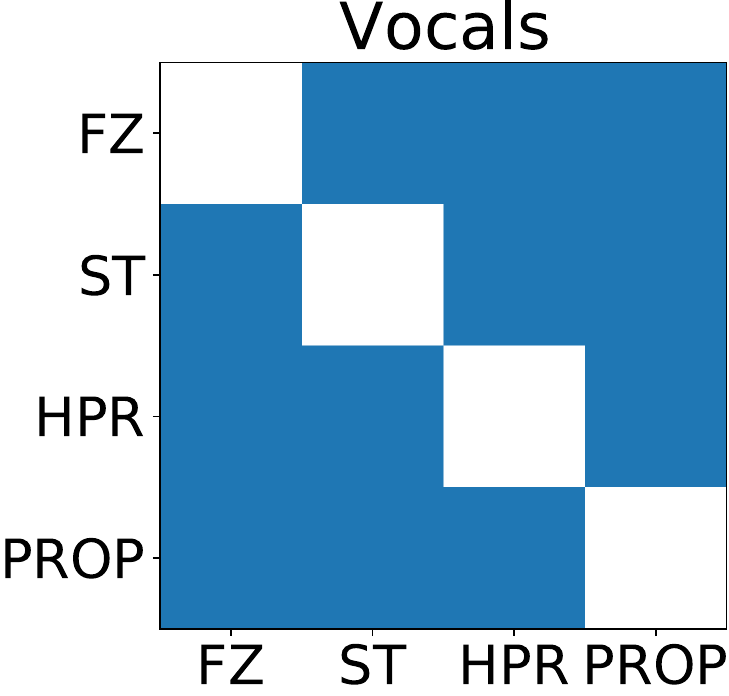}
		\caption{}
		\vspace*{4mm}
	\end{subfigure}
	\caption{Thresholded $p$--values resulting from the Wilcoxon signed-rank test over sinusoidal separation data. Statistical significance (p $ \leq \alpha, \alpha = 0.05$) is highlighted by coloring the cell.}\label{fig:WilTones}
\end{figure*}
\begin{figure*}	[t!]
	\centering
	\begin{subfigure}[t]{0.15\textwidth}
		\centering
		\includegraphics[width=\hsize]{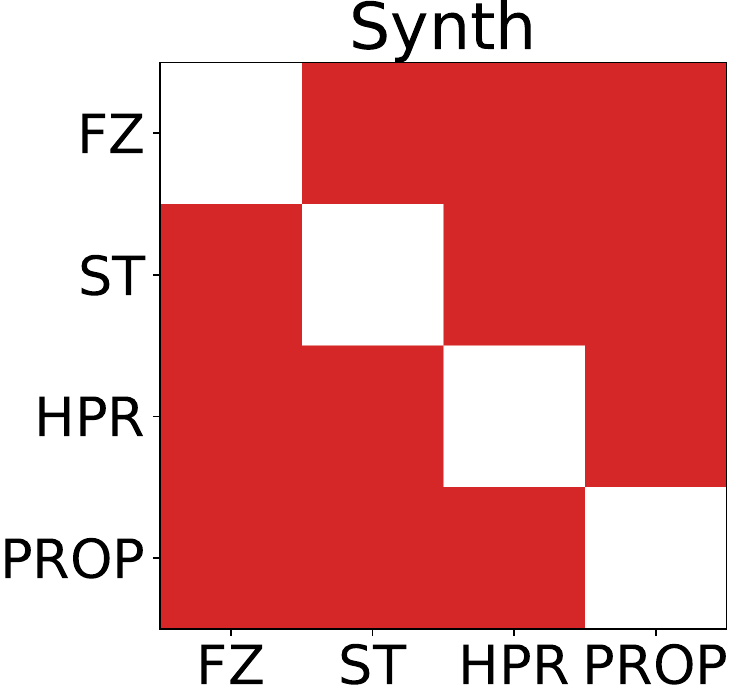}
		\caption{}
		\vspace*{4mm}
	\end{subfigure}
	\hspace{0.03\textwidth}
		\begin{subfigure}[t]{0.15\textwidth}
		\centering
		\includegraphics[width=\hsize]{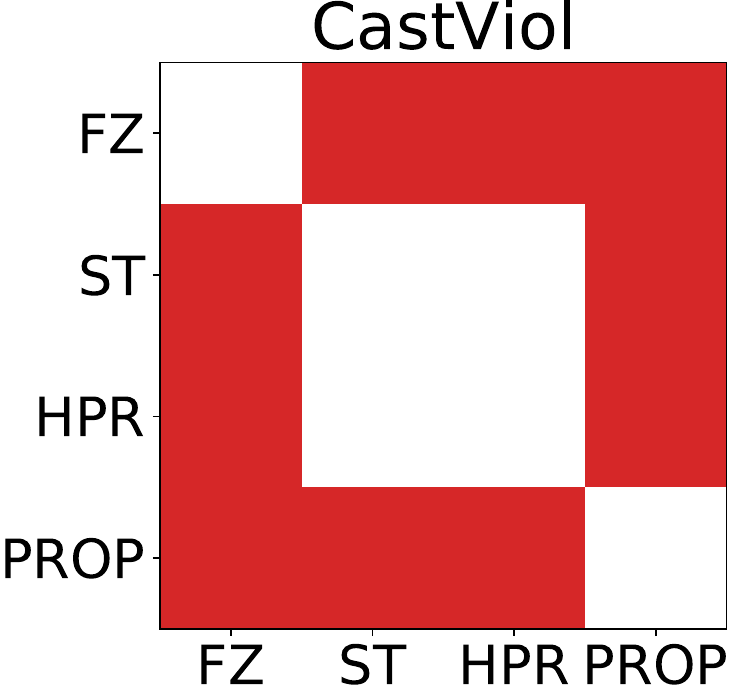}
		\caption{}
		\vspace*{4mm}
	\end{subfigure}
	\hspace{0.03\textwidth}
		\begin{subfigure}[t]{0.15\textwidth}
		\centering
		\includegraphics[width=\hsize]{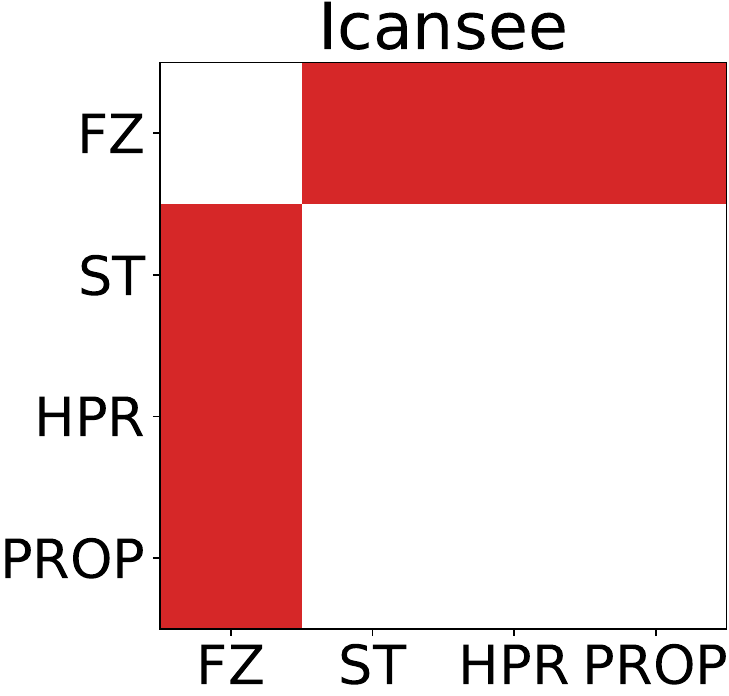}
		\caption{}
		\vspace*{4mm}
	\end{subfigure}
	\hspace{0.03\textwidth}
		\begin{subfigure}[t]{0.15\textwidth}
		\centering
		\includegraphics[width=\hsize]{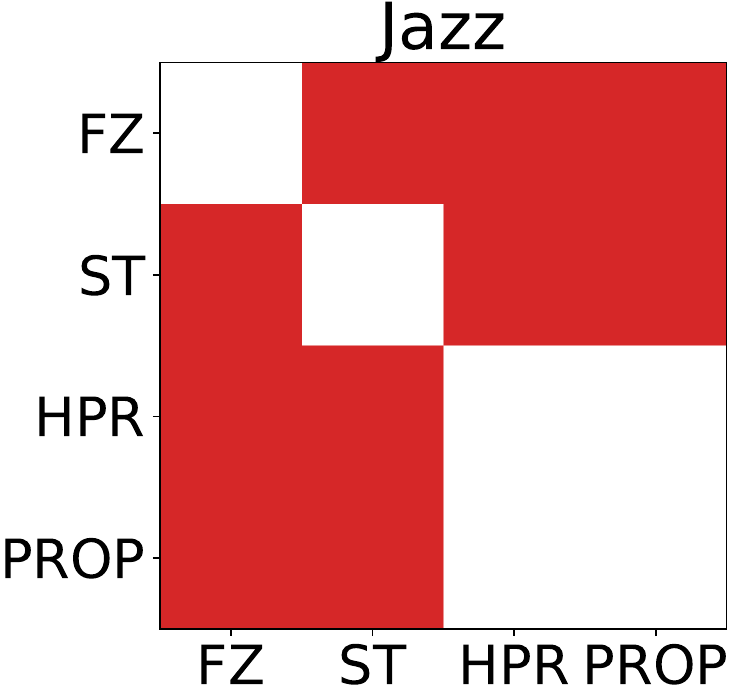}
		\caption{}
		\vspace*{4mm}
	\end{subfigure}
	\hspace{0.03\textwidth}
	\begin{subfigure}[t]{0.15\textwidth}
		\centering
		\includegraphics[width=\hsize]{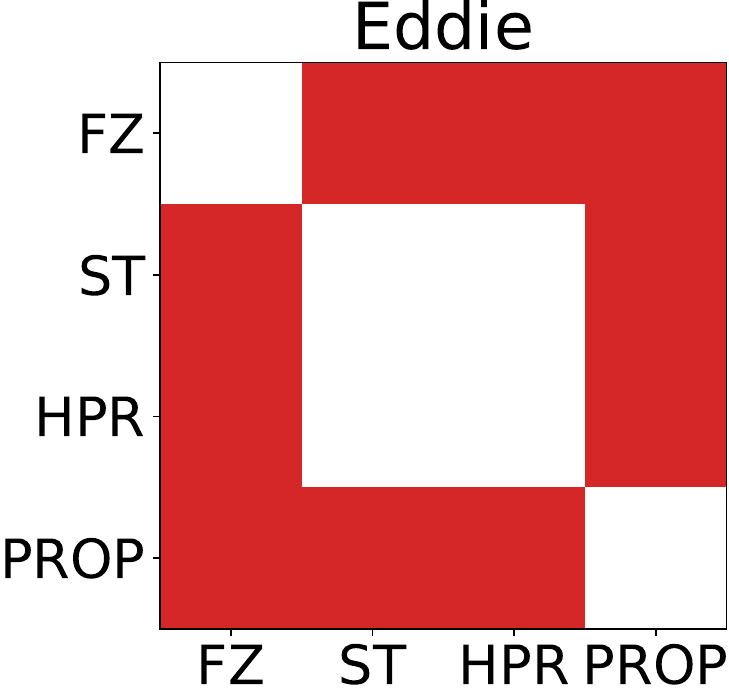}
		\caption{}
		\vspace*{4mm}
	\end{subfigure}
	\hspace{0.03\textwidth}
		\begin{subfigure}[t]{0.15\textwidth}
		\centering
		\includegraphics[width=\hsize]{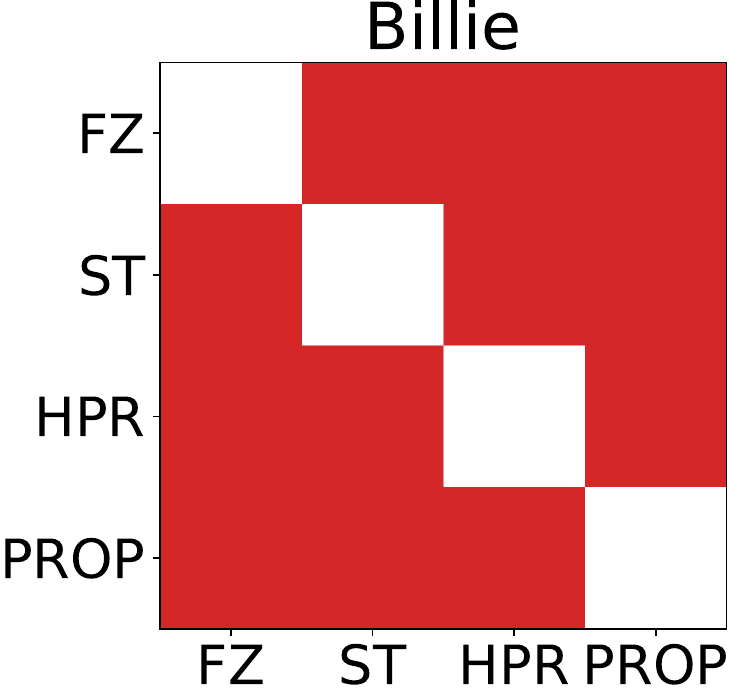}
		\caption{}
	\end{subfigure}
	\hspace{0.03\textwidth}
		\begin{subfigure}[t]{0.15\textwidth}
		\centering
		\includegraphics[width=\hsize]{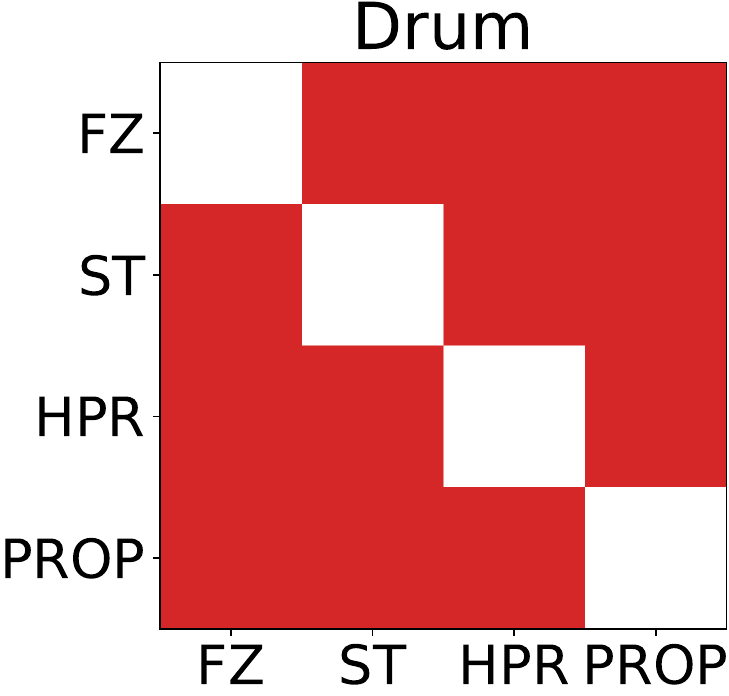}
		\caption{}
	\end{subfigure}
	\hspace{0.03\textwidth}
	\begin{subfigure}[t]{0.15\textwidth}
		\centering
		\includegraphics[width=\hsize]{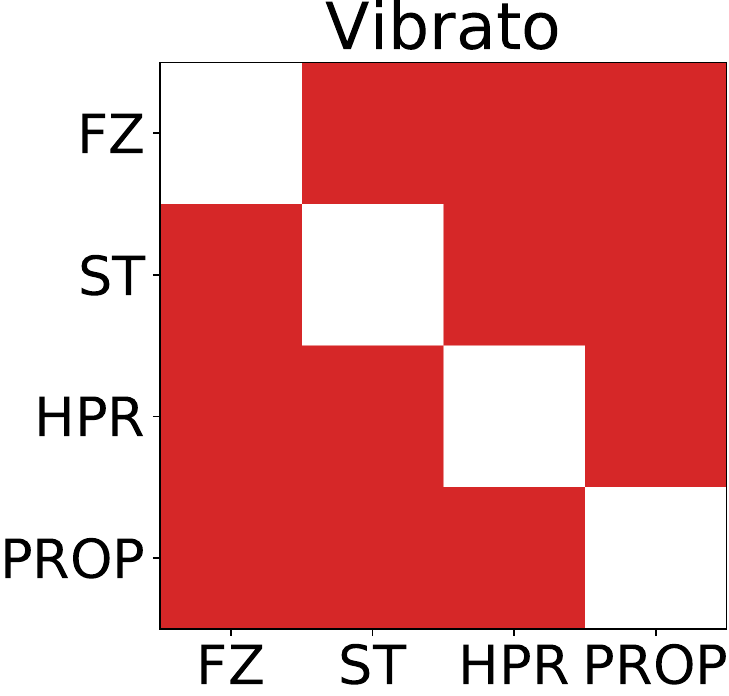}
		\caption{}
	\end{subfigure}
	\hspace{0.03\textwidth}
	\begin{subfigure}[t]{0.15\textwidth}
		\centering
		\includegraphics[width=\hsize]{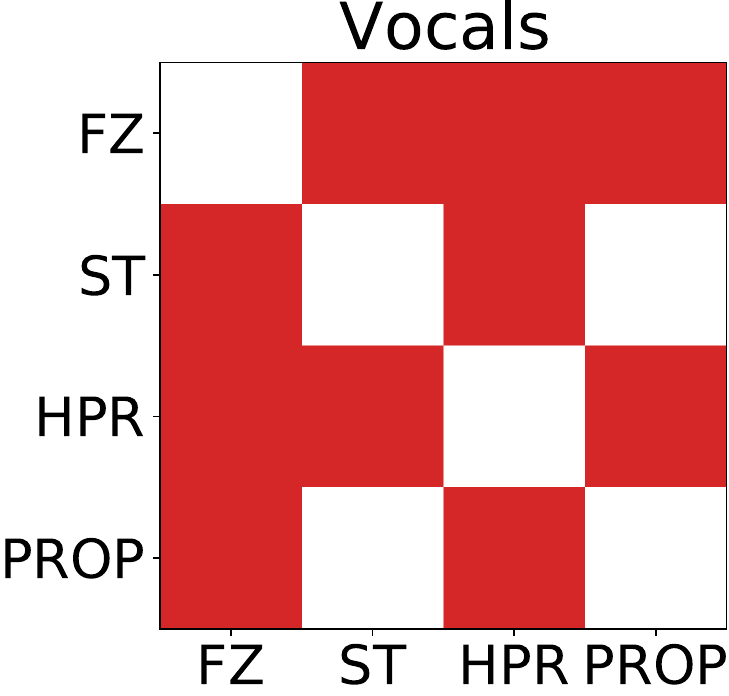}
		\caption{}
		\vspace*{4mm}
	\end{subfigure}
	\caption{Thresholded $p$--values of transient separation data, cf.~Fig.~\ref{fig:WilTones}.  Statistical significance (p $ \leq \alpha, \alpha = 0.05$) is highlighted by coloring the cell.}\label{fig:WilTransient}
\end{figure*}

Mean Opinion Scores (MOS) were computed from the ratings given by the subjects to estimate the quality of the decomposition methods under test. Boxplots displaying the distribution of the data for the sines and transients extraction are shown, respectively, in Figs.~\ref{fig:BoxSines} and \ref{fig:BoxTransient}. Data distribution can also be observed via histograms, available at the companion webpage \cite{webpage}.

Overall, the proposed method consistently performed better than or as well as HPR for every audio excerpt, for both sinusoidal and transient decompositions. The proposed methods achieved a median MOS score between 3.0 and 5.0 (``good'' to ``excellent'') in all but one test case, see Figs.~\ref{fig:BoxSines} and \ref{fig:BoxTransient}. The HPR method was below 3.0 in two cases. The ST method generally scored intermediate values, with the exception of \textit{Vibrato}, where ST outperformed the other separation methods: this was expected, considering that ST was designed purposely for sound mixtures presenting vibrato. FZ was the lowest-ranked algorithm with a significant difference in the subjective ratings from the other three methods, as appears from Figs.~\ref{fig:BoxSines} and \ref{fig:BoxTransient}.

Considering the sines extraction only, the proposed method performed quite similarly to HPR in Fig.~\ref{fig:BoxSines} and significantly better only for the \textit{Vocals} and \textit{Drum} excerpts, where the fuzzy soft masking helped in the preservation of the tonality variations. However, the median value for the proposed method was consistently higher than that of the HPR method.

A larger improvement was observed in the transient decomposition in Fig.~\ref{fig:BoxTransient}. Considering the median of the distributions, the proposed method surpassed HPR for all excerpts but \textit{Icansee} and \textit{Jazz}. ST proved to be a competitive algorithm for \textit{Vocals}.
The large amount of variance in the data came from the absence of a proper separation ``reference'', i.e.~an upper limit for the subjective grading in each trial. The subjects had to apply their own scale during the grading process, which consequently lead to data which are distributed in a non--Gaussian fashion. This was also confirmed by an inspection of data skewness and kurtosis, which is reported in the companion website \cite{webpage}. 

Results were validated by conducting further statistical analysis and determining whether there was statistical significance in the data distribution, i.e.~the difference between the distribution of data for different methods had statistical significance. In this case, the observation of non--Gaussian distributions called for a non--parametric paired difference test. The Wilcoxon signed--rank test was chosen for the task. For a 95\% confidence interval, statistical significance was achieved if the signed--rank test returned a $p$-value below the threshold $\alpha = 0.05$. Thresholded resulting $p$–values for sinusoidal and transient separation data are shown, respectively, in Figs.~\ref{fig:WilTones} and \ref{fig:WilTransient}. 

The results showed that statistical significance for the difference in data distribution from the proposed and the competitive methods was achieved for at least one of the two components (sinusoidal or transient separation) for seven excerpts out of nine, with the transient separation being the discriminant factor in six cases out of seven. \textit{Icansee} and \textit{Jazz} were the two most complex mixtures among the collected audio samples: this hinted that the more complex the sound mixture, the harder it became to discriminate the separation performance. The traditional statistical analysis via ANOVA and paired $t$--test, carrying similar results, is reported in the companion website \cite{webpage}.

\begin{figure}[t!]
\center
\begin{subfigure}[t]{0.75\columnwidth}
\centering
\includegraphics[width=\textwidth]{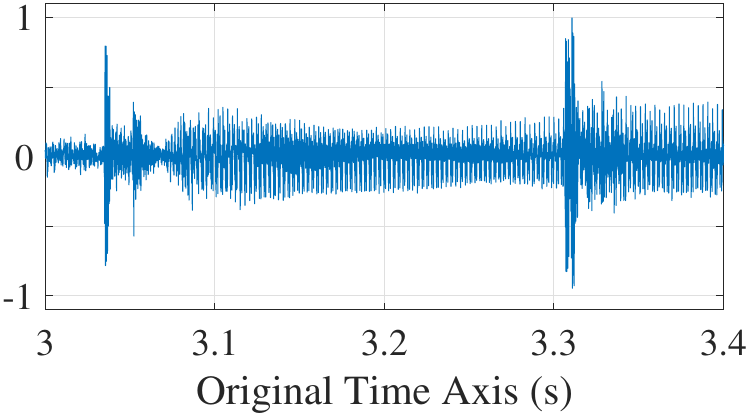}
\caption{}
\end{subfigure}
\begin{subfigure}[t]{0.75\columnwidth}
\centering
\includegraphics[width=\textwidth]{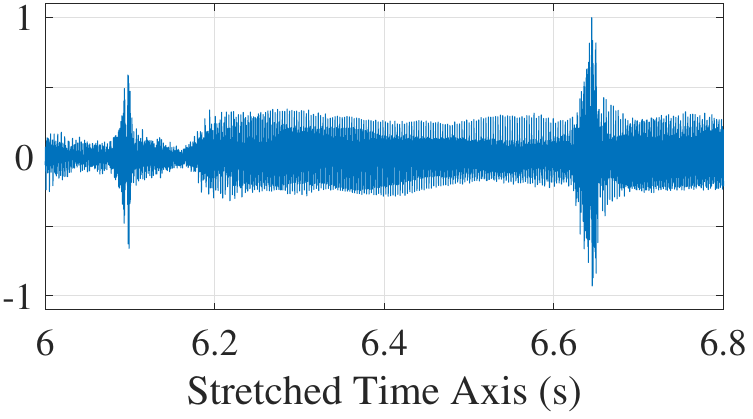}
\caption{}
\end{subfigure}
\begin{subfigure}[t]{0.75\columnwidth}
\centering
\includegraphics[width=\textwidth]{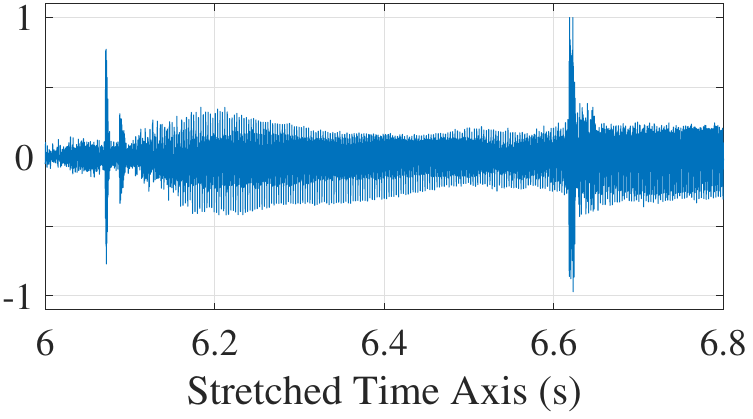}
\caption{}
\end{subfigure}
\caption{\label{fig:TSM} Comparison between (a) a section of \textit{CastViol} and its modifications via (b) fuzzy phase vocoder (FPV) \cite{damskagg2017audio} and (c) its STN enhancement (PROP), for a TSM factor of 2. Note the different time scales in the subfigures.}
\end{figure}

\section{APPLICATION TO TIME STRETCHING}
\label{sec:TSM}

The proposed method is adapted to audio time-scale modification (TSM), which is a suitable application for the STN decompositions \cite{damskagg2017audio,fierro2020towards,driedger2014tsm}. For this purpose, the fuzzy phase vocoder developed by Damskägg and Välimäki \cite{damskagg2017audio}, which received the highest average score in a recent comparison of audio time-stretching methods \cite{roberts2021}, is modified to include the proposed decomposition method. The refined separation allows for the transients to be preserved and repositioned onto the stretched time axis \cite{nagel2009novel}, while the sinusoidal and noise components are processed via the phase vocoder with identity phase locking and phase randomization, respectively. 

The enhanced TSM processing for a section of the \textit{CastViol} excerpt that has been slowed down to half speed is shown in Fig.~\ref{fig:TSM}. The sound processed with the original fuzzy phase vocoder visibly suffers from transients smearing, a recurring phenomenon in phase-vocoder-based TSM \cite{robel2003new}, With the STN-enhanced version, the transients appear much sharper and better resemble the ones visible in the original signal.


\begin{figure}[t!]
\center
\begin{subfigure}[t]{\columnwidth}
\centering
\includegraphics[width=\textwidth]{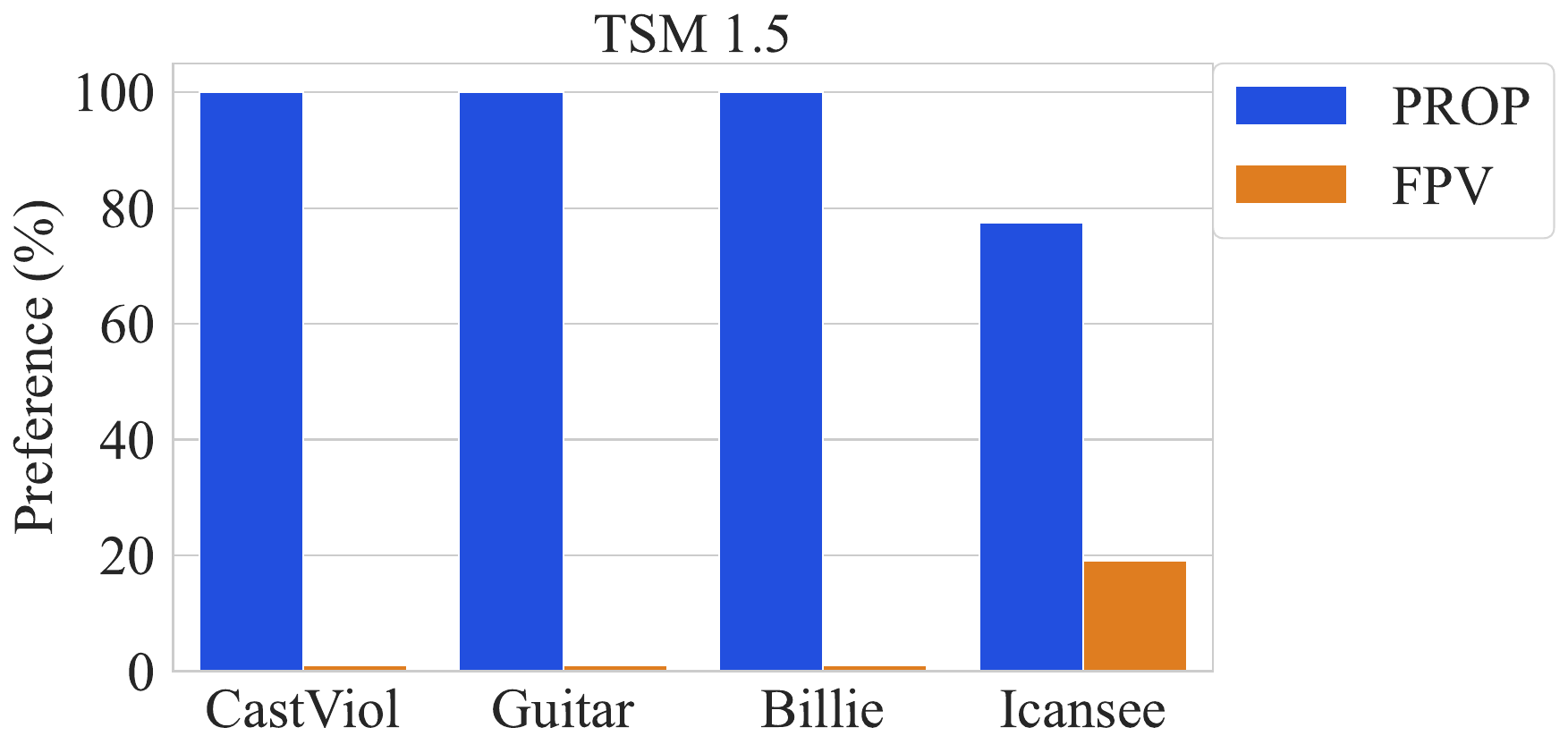}
\caption{}
\end{subfigure}
\begin{subfigure}[t]{\columnwidth}
\centering
\includegraphics[width=\textwidth]{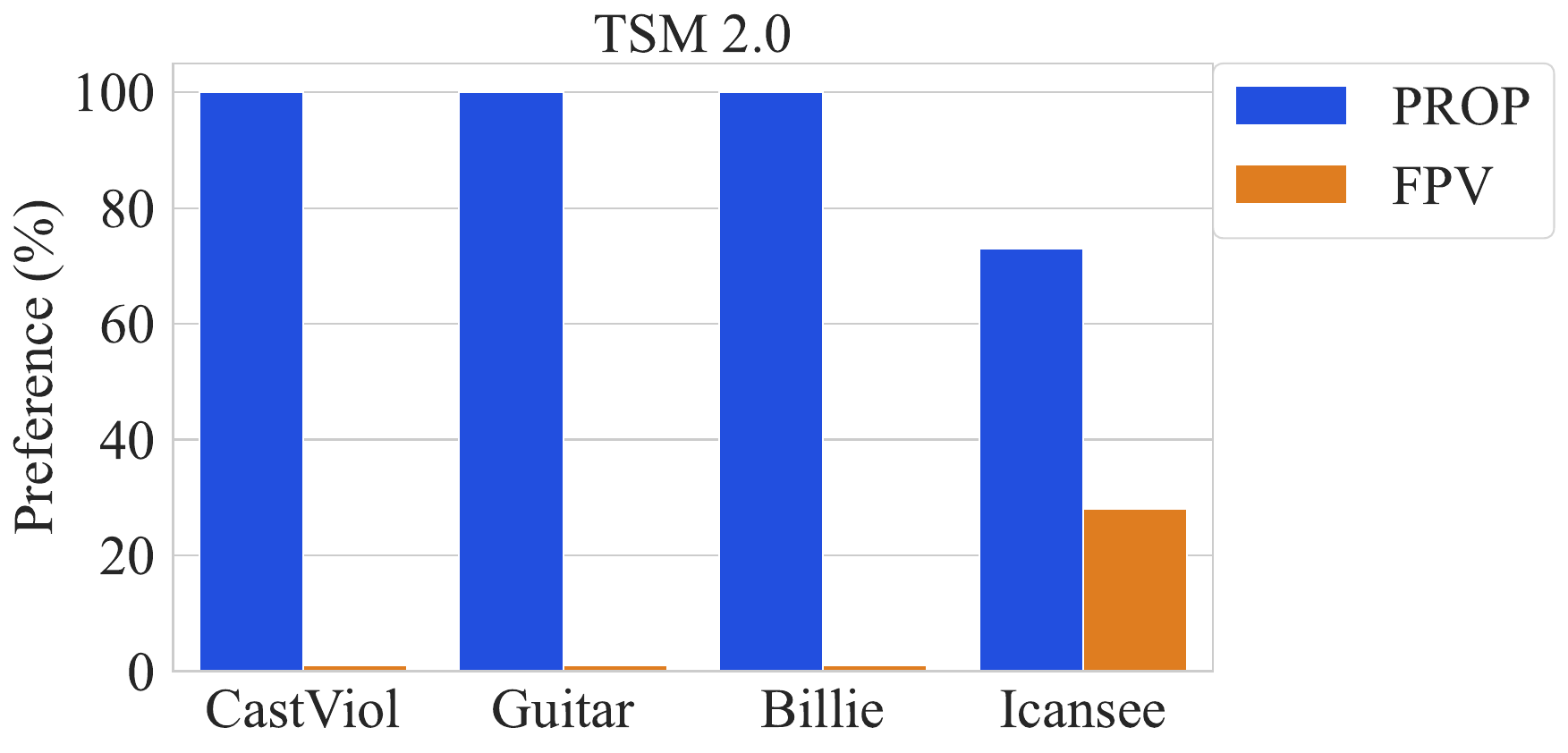}
\caption{}
\end{subfigure}
\caption{\label{fig:barplots} Subject preferences between original fuzzy phase vocoder (FPV) \cite{damskagg2017audio} and the STN modification (PROP), for TSM factors (a) 1.5 and (b) 2.}
\end{figure}

\subsection{Comparison}

A preference test was conducted to compare the original fuzzy phase vocoder (FPV) \cite{damskagg2017audio} with the proposed STN-modified one (PROP), in order to evaluate the enhancement brought by STN decomposition to a suitable method. Eleven experienced listeners participated in the test, which was realized on the same hardware and with similar modalities as the one described in Sec.~\ref{sec:testdesign}. Subjects were asked to listen to a reference sound and to select their preferred time-stretched version in terms of sound quality from two options, processed respectively with FPV and PROP. Four audio excerpts (\textit{CastViol, Billie}, and \textit{Icansee} from the previous test, plus a \textit{Guitar} plucking sound) were time-stretched with factors 1.5 and 2. Loudness normalization was applied to compensate for the non-perfect reconstruction of FPV. Results are visualized in Fig.~\ref{fig:barplots}, and all test sounds are available on the companion webpage \cite{webpage}.

Test subjects showed a strong preference for PROP over all samples for both time-stretching factors. A minority expressed a preference for FPV for \textit{Icansee}, mentioning that while the transients' ``punch" was well retained by the other method (PROP), that created a dissonance with the noisy part of the transient, which goes through the phase vocoder and phase randomization and is heavily smeared. This hints that a different processing for the noise component, which can now be isolated through the STN decomposition, could further improve the audio time-stretching performance.

\section{CONCLUSION}
\label{sec:conclusion}

In this paper, the three-way sound decomposition into sines, transients, and noise using fuzzy logic was enhanced. A set of soft spectral masks was derived to fulfill the task while preserving the perfect reconstruction property. Using such soft masks, the novel two–stage STN decomposition method proposed in this paper allows a single spectral bin to be simultaneously classified either as sine and noise, or as transient and noise. Soft masking positively affects the decomposition by attenuating or removing common artifacts, e.g. musical noise or loss of transient presence.

The results of a subjective listening test against three other methods showed that the proposed decomposition method typically improves the separation quality in terms of transient extraction, with a comparable performance for sines extraction with the previous best method. It was also shown how the complexity of the audio signal affects the quality of the decomposition. For instance, the proposed separation method struggles when the sinusoidal part contains vibrato, as does the competing previous method. 

The proposed method can help improve sound quality in many audio processing tasks. A successful application to audio time stretching was shown to improve the performance of the state-of-the-art algorithm.


\section{ACKNOWLEDGMENTS}
This work belongs to the activities of the ``Nordic Sound and Music Computing Network---NordicSMC'', NordForsk project number 86892. The work of Leonardo Fierro was funded by the Aalto ELEC Doctoral School. The authors are grateful to Dennis Bontempi for the helpful discussions, and to Alec Wright for proofreading.


\normalsize
\bibliography{mybib.bib}




\end{document}